\documentclass[12pt]{article}

\usepackage{graphics,cite,amssymb,epsfig,float,psfrag}
\usepackage{rotating}
\newcommand{\lsim}{\raisebox{-0.13cm}{~\shortstack{$<$ \\[-0.07cm] $\sim$}}~} 
\newcommand{\gsim}{\raisebox{-0.13cm}{~\shortstack{$>$ \\[-0.07cm] $\sim$}}~} 
\newcommand{\beq}{\begin{eqnarray}} 
\newcommand{\eeq}{\end{eqnarray}} 
\newcommand{\tb}{\tan \beta}
\newcommand{\s}{\\ \vspace*{-4mm}}

\begin{document}

\vspace{.8cm}


 
\vspace*{1.4cm}

\begin{center}

{\large\bf The post-Higgs MSSM scenario$\,$:}

\vspace*{.3cm}

{\large\bf Habemus MSSM?}

\vspace*{.7cm}

{\sc A. Djouadi$^1$, L. Maiani$^{2,3}$, G. Moreau$^1$,  
A. Polosa$^2$,} \

\vspace*{.1cm}

{\sc J. Quevillon$^1$} and {\sc V. Riquer$^2$}

\vspace*{.7cm}

\begin{small}

$^1$ Laboratoire de Physique Th\'eorique, Universit\'e Paris--Sud and  CNRS,  

F--91405 Orsay, France.

\vspace*{.1cm}

$^2$ Department of Physics and INFN, ``Sapienza" Universt\`a di Roma,  

Pizzale Aldo Moro 5, I--00185 Roma, Italia.

\vspace*{.1cm}

$^3$ Theory Unit, CERN, 1211 Gen\`eve 23, Switzerland.

\end{small}

\end{center}

\vspace*{1cm}

\begin{abstract} 

We analyze the Minimal Supersymmetric extension of the Standard Model that we 
have after the discovery of the Higgs boson at the LHC, the hMSSM (habemus MSSM?),  
i.e. a model in which the lighter $h$ boson has a mass of approximately
125  GeV  which, together with the non-observation of superparticles at the LHC,
indicates  that the  SUSY--breaking scale $M_S$ is rather high, $M_S \gsim 1$
TeV.   We first  demonstrate  that the value $M_h \approx 125$ GeV  fixes the
dominant radiative corrections that enter  the MSSM Higgs boson masses, leading
to a Higgs sector that can be described, to a  good approximation,  by only
two free parameters. In a second step, we consider the direct supersymmetric 
radiative corrections and show that, to a good approximation, the phenomenology  
of the lighter Higgs state can be described  by its mass and 
three couplings:  those to massive gauge bosons and to top and bottom quarks. We
perform a fit of these couplings using the latest LHC data on the production
and decay rates of the light $h$ boson and combine it with the limits from the negative search of the heavier $H,A$ and $H^\pm$ states, taking into account the current
uncertainties.

\end{abstract} 

\thispagestyle{empty}

\newpage
\setcounter{page}{1}

\subsection*{1. Introduction} 

The observation at the LHC of a Higgs particle  with a mass of 125 GeV
\cite{LHC-Higgs} has important  implications for Supersymmetric (SUSY) and, 
in particular, for the Minimal Supersymmetric Standard Model (MSSM). In this
extension, the Higgs sector consists of two scalar doublet fields $H_u$ and
$H_d$  that lead, after electroweak symmetry breaking, to five Higgs states, two
CP--even $h$ and $H$, a CP--odd $A$ and two charged $H^\pm$ bosons
\cite{Review,Mh-max}. At tree level, the masses of these particles and their
mixings are described  by only two parameters usually chosen to be the  ratio of
the vacuum expectations values of the two doublet fields $\tb\!=\!v_d/v_u$  and
the mass $M_A$ of the pseudoscalar Higgs boson. However, as is well known, the
radiative corrections play a very important role  as their dominant component
grows like  the fourth power of the top quark mass, logarithmically with the
supersymmetry breaking scale $M_S$  and quadratically with the stop mixing
parameter $A_t$; see e.g. Refs.~\cite{Mh-max,CR-1loop,CR-eff}.  

The impact of the Higgs discovery is two--fold. On the one hand, it gives
support to the MSSM  in which the lightest Higgs boson is predicted to have a
mass below $\approx 130$ GeV when the radiative corrections  are included
\cite{Mh-max,CR-1loop,CR-eff}. On the other hand, the fact that the measured
value $M_h \approx 125$ GeV is close to this upper mass limit implies that the
SUSY--breaking scale $M_S$ might be rather high. This is backed up by the
presently strong limits on supersymmetric particle masses from direct searches
that indicate that the SUSY partners of the strongly interacting particles, the
squarks and gluinos, are heavier than $\approx 1$ TeV  \cite{LHC-SUSY}. Hence,
the MSSM that we currently have, and that we call hMSSM (habemus MSSM?) in the
subsequent discussion,  appears to have $M_h \approx 125$ GeV  and $M_S \gsim 1$
TeV. 

It was pointed out in Refs.~\cite{R1,R1p,O1} that when the information 
$M_h\!=\!125$ GeV is taken into account, the MSSM Higgs  sector with  solely the
dominant radiative correction to the Higgs boson masses included, can be again
described with only the two free parameters $\tb$  and $M_A$ as it was the case
at tree--level. In other words,  the dominant radiative corrections that involve
the SUSY parameters  are fixed by the value of $M_h$. In this paper, we show
that to a good approximation, this remains true even  when  the full set of
radiative corrections to the Higgs masses at the two--loop level is included.
This is demonstrated in particular by  performing a full scan on the MSSM
parameters that have an impact on the Higgs sector such as for instance $\tb$ and 
the stop and sbottom mass and mixing parameters. The subleading radiative corrections 
are shown to have little impact on the mass and mixing of the heavier
Higgs bosons when these SUSY parameters are varied in a reasonable range. 

Nevertheless, there are also possibly large direct SUSY  radiative corrections
that modify the  Higgs boson couplings and  which might alter this simple
picture. Among such corrections are, for instance, the stop contribution 
\cite{Stop,Stop2} to the dominant Higgs production mechanism at the LHC, the gluon
fusion process $gg\to h$, and to the important decay into two photons $h\to
\gamma \gamma$, and the additional one--loop vertex corrections to the $h$
couplings to $b$--quarks that grow with $\tb$ \cite{Deltab}. In the most general
case, besides $M_h$, seven couplings  need to be  considered to fully describe the
properties of the  observed $h$ boson: those to gluons, photons, massive gauge
bosons,  $t,b,c$ quarks and $\tau$ leptons. However, we show that  given the
accuracy that is foreseen at the LHC, a good approximation is to consider the
three effective couplings to $t,b$  quarks and to $V=W/Z$ bosons, $c_t,c_b$ and
$c_V$, as it was suggested in Ref.~\cite{R2}. Following the approach
of Ref.~\cite{O2} for the  inclusion of the current theoretical and experimental
uncertainties, we perform a fit of these three couplings using  the latest LHC
data on the production and decay rates of the lighter $h$ boson and the limits
from the negative search of the heavier $H,A$ and $H^\pm$ MSSM states. 

The best fit points to low values of $\tb$ and  to $M_A$ values of the order 
of 500 GeV, leading to a spectrum in the Higgs sector that can be fully
explored at the 14 TeV LHC.

Almost one year after the Higgs discovery at the LHC,  these two aspects will be 
discussed in
the next two sections. A brief discussion and a conclusion are given in section 4 
and a short Appendix collects a set of formulae used in this analysis.

\subsection*{2. Post Higgs discovery parametrisation of radiative corrections}

In the MSSM, the tree--level masses of the CP--even $h$ and $H$  bosons depend
on $M_A$, $\tb$ and the $Z$ boson mass. However, many parameters of the MSSM
such as the SUSY scale,  taken to be the geometric average of the stop masses
$M_S= \sqrt {m_{\tilde t_1} m_{\tilde t_2} }$, the stop/sbottom trilinear
couplings $A_{t/b}$ or the higgsino mass  $\mu$  enter $M_h$ and $M_H$ through
radiative corrections.  In the basis $(H_d,H_u)$, the CP--even Higgs  mass
matrix  can be written as:
\beq
M_{S}^2=M_{Z}^2
\left(
\begin{array}{cc}
  c^2_\beta & -s_\beta c_\beta \\
 -s_\beta c_\beta & s^2_\beta \\
\end{array}
\right)
+M_{A}^2
\left(
\begin{array}{cc}
 s^2_\beta & -s_\beta c_\beta \\
 -s_\beta c_\beta& c^2_\beta \\
\end{array}
\right)
+
\left(
\begin{array}{cc}
 \Delta {\cal M}_{11}^2 &  \Delta {\cal M}_{12}^2 \\
 \Delta {\cal M}_{12}^2 &\Delta {\cal M}_{22}^2 \\
\end{array}
\right)
\eeq

where we use the short--hand notation $s_\beta \equiv \sin\beta$ etc$\dots$  and
have introduced the radiative corrections  by a $2\times 2$ general matrix  
$\Delta {\cal M}_{ij}^2$. One can then easily derive the neutral CP even Higgs 
boson masses and the mixing angle $\alpha$ that diagonalises the $h,H$ 
states\footnote{A   different definition for the mixing angle $\alpha$, namely
$\alpha \to \frac{\pi}{2}- \alpha$, has been adopted in
Refs.~\cite{R1,R1p,R2}.}, $H= \cos\alpha H_d^0 + \sin\alpha H_u^0$  and $h=
-\sin\alpha H_d^0 + \cos\alpha H_u^0$ 
\begin{eqnarray}
\hspace{-1.0cm}
M_{h/H}^2&=&\frac{1}{2} \big( M_{A}^2+M_{Z}^2+ \Delta {\cal M}_{11}^2+ 
\Delta {\cal M}_{22}^2  \mp  \sqrt{ M_{A}^4+M_{Z}^4-2 M_{A}^2 M_{Z}^2 
c_{4\beta} +C} \big) \\
\hspace{-1.0cm}
\tan \alpha&=&\frac{2\Delta {\cal M}_{12}^2 - (M_{A}^2 + M_{Z}^2) s_{\beta}}
{ \Delta {\cal M}_{11}^2 -  \Delta {\cal M}_{22}^2 + (M_{Z}^2-M_{A}^2)
c_{2\beta} + 
\sqrt{M_{A}^4 + M_{Z}^4 - 2 M_{A}^2 M_{Z}^2 c_{4\beta} + C}}
\end{eqnarray}
\vspace*{-5mm}
\begin{eqnarray}
C=  4 \Delta {\cal M}_{12}^4\! + \!( \Delta {\cal M}_{11}^2 \!- \! 
\Delta {\cal M}_{22}^2)^2 \!- \! 
 2 (M_{A}^2 \! - \! M_{Z}^2)( \Delta {\cal M}_{11}^2 \! - \! \Delta M_{22}^2) 
 c_{2\beta} \!   - \!
 4 (M_{A}^2 \! + \!M_{Z}^2)  \Delta {\cal M}_{12}^2 s_{2\beta} \nonumber
\end{eqnarray}

In previous analyses \cite{R1,R1p,O1}, we have  assumed that in the $2\times 2$
matrix for the radiative corrections,  only the $\Delta{\cal M}^{2}_{22}$ entry
which involves the by far dominant stop--top sector correction, is relevant, 
$\Delta{\cal M}^{2}_{22} \gg \Delta{\cal M}^{2}_{11}, \Delta{\cal M}^{2}_{12}$.
This occurs, for instance, in the so--called $\epsilon$ approximation
\cite{CR-1loop}  and its refinements \cite{CR-eff} that are given in eqs.~(A2)
and (A3) of the Appendix. In this case,  one can simply trade $\Delta {\cal
M}^{2}_{22}$ for the by now known $M_h$ using  
\beq
\Delta {\cal M}^{2}_{22}= \frac{M_{h}^2(M_{A}^2  + M_{Z}^2 -M_{h}^2) - M_{A}^2 M_{Z}^2 c^{2}_{2\beta} } { M_{Z}^2 c^{2}_{\beta}  +M_{A}^2 s^{2}_{\beta} -M_{h}^2}
\eeq
In this case, one can simply write  $M_H$ and $\alpha$ in terms of 
$M_A,\tb$ and $M_{h}$:
\begin{eqnarray}
{\rm hMSSM}:~~ 
\begin{array}{l} 
M_{H}^2 = \frac{(M_{A}^2+M_{Z}^2-M_{h}^2)(M_{Z}^2 c^{2}_{\beta}+M_{A}^2
s^{2}_{\beta}) - M_{A}^2 M_{Z}^2 c^{2}_{2\beta} } {M_{Z}^2 c^{2}_{\beta}+M_{A}^2
s^{2}_{\beta} - M_{h}^2} \\
\ \ \  \alpha = -\arctan\left(\frac{ (M_{Z}^2+M_{A}^2) c_{\beta} s_{\beta}} {M_{Z}^2
c^{2}_{\beta}+M_{A}^2 s^{2}_{\beta} - M_{h}^2}\right)
\end{array}
\label{wide} 
\end{eqnarray}
In this section, we will check the validity of the $ \Delta{\cal M}^{2}_{11}=
\Delta{\cal M}^{2}_{12}=0$ approximation. To do so,  we first consider the 
radiative corrections when the subleading contributions  proportional to $\mu,
A_t$ or $A_b$ are included in the form of eqs.~(A4--A6) of the  Appendix,
that is expected to be a good approximation \cite{Mh-max,Pietro},  and in which one has $
\Delta{\cal M}^{2}_{11} \neq \Delta{\cal M}^{2}_{12} \neq 0$.

\begin{figure}[!h]
\vspace*{-2mm}
\begin{center} 
\begin{tabular}{c}
\includegraphics[scale=0.67]{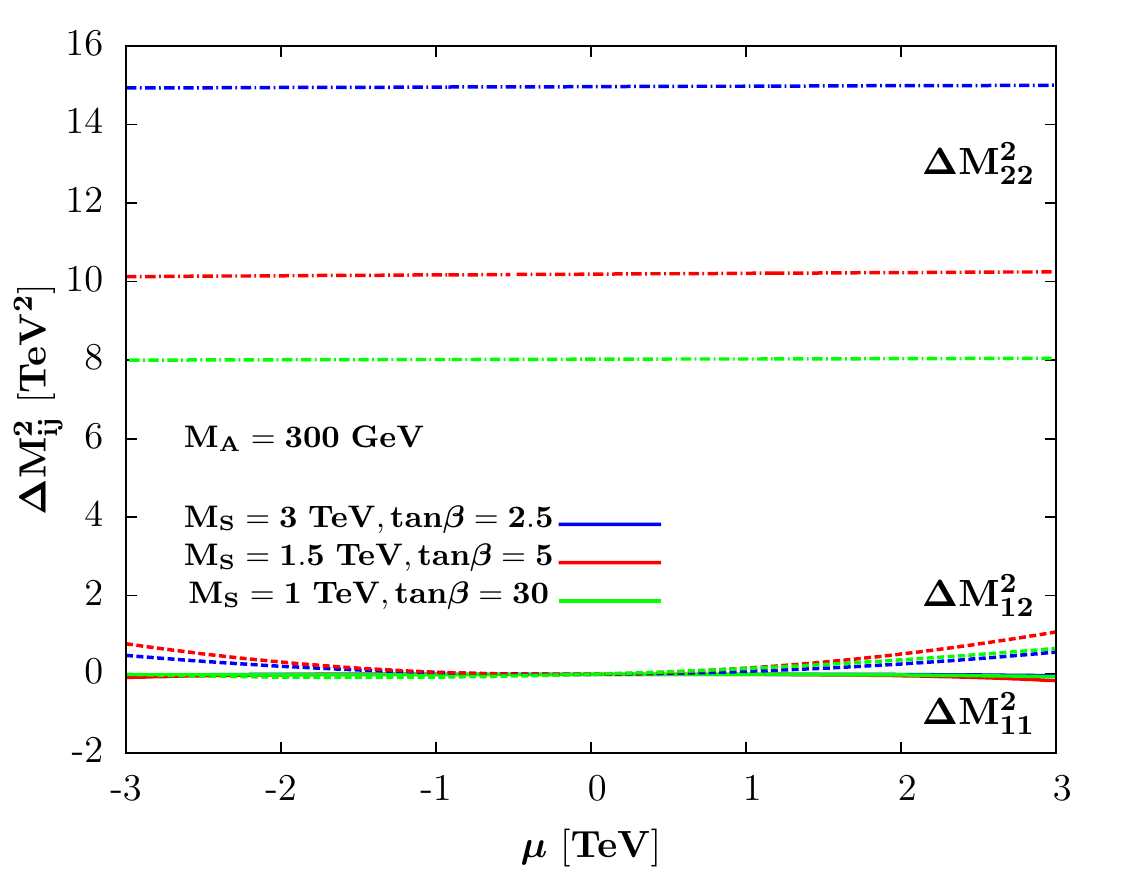} \\
\end{tabular}
\vspace*{-2mm}
\caption{{\small
The entries $\Delta {\cal M}^{2}_{11}$ (solid), $\Delta {\cal M}^{2}_{12}$
(dashed), and $\Delta {\cal M}^{2}_{22}$ (dotted-dashed lines) of the  radiative
corrections matrix as  functions of $\mu$ with a fixed  $M_A\!=\!300$~GeV for three
different  $(M_S, \tb)$ sets and $A_{t}$  such that it accommodates the mass
range $M_h=123$--129~GeV.}}
\label{approx0}
\end{center}
\vspace*{-6mm}
\end{figure}

As a first step we only consider the stop-top sector corrections which enter the
$\Delta {\cal M}^{2}_{ij}$ terms and confront in Fig.~\ref{approx0}, the  values
of $\Delta {\cal M}^{2}_{11}$, $\Delta {\cal M}^{2}_{12}$ to $\Delta {\cal
M}^{2}_{22}$  for three different scenarios with  $M_A\! = \! 300$~GeV (i.e. before the
onset of the decoupling regime $M_A\! \gg \! M_Z)$:  $M_S\!=\!3$~TeV and $\tb\!=\!2.5$, $M_S\! =\! 1.5$~TeV
and $\tb\! =\! 5$, $M_S\! =\! 1$~TeV and $\tb\! =\! 30$.  The  parameter $A_{t}$ is adjusted  in
order to accommodate a light Higgs boson with a mass $M_h=126\pm 3$ GeV,
including  an expected  theoretical and experimental uncertainty of 3 GeV
\cite{error}. One observes that for reasonable  $\mu$ values,  one obtains
naturally $\Delta {\cal M}^{2}_{11},\Delta {\cal M}^{2}_{12} \ll \Delta {\cal
M}^{2}_{22}$.

We have verified that the situation is not very different if the corrections in
the sbottom sector are also included: assuming $A_b=A_t$,  we also obtain the
hierarchy $\Delta {\cal M}^{2}_{11},\Delta {\cal M}^{2}_{12} \ll \Delta {\cal
M}^{2}_{22}$  for $\mu \lsim 3$ TeV  even for $\tb=30$ where contributions
$\propto \mu \tb$ become important.

\begin{figure}[!h]
\vspace*{-2mm}
\begin{center}
\begin{tabular}{cc}
\includegraphics[scale=0.67]{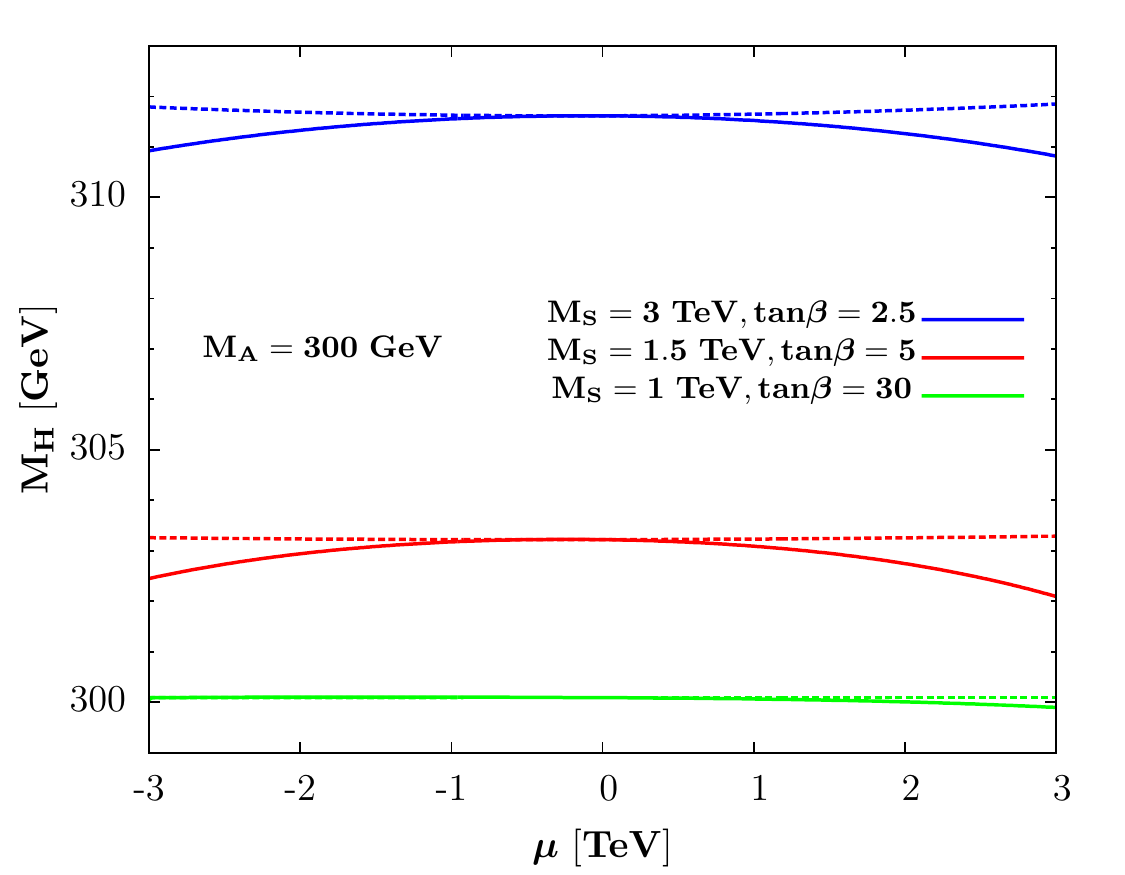} &
\includegraphics[scale=0.67]{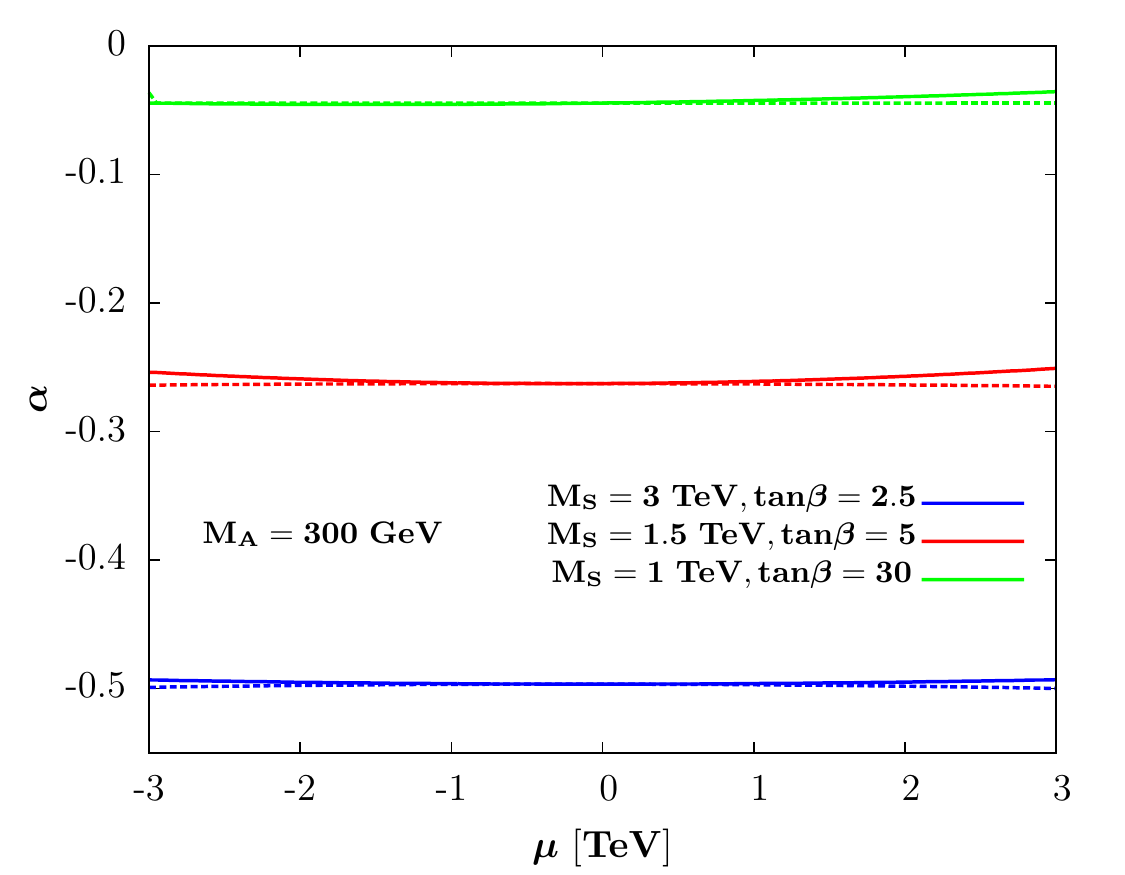}
\end{tabular}
\vspace*{-5mm}
\caption{\small
The mass of the heavier CP--even $H$ boson (left) and the mixing angle $\alpha$ 
(right) as a function of $\mu$  with (solid lines) and without (dashed) the
off--diagonal components components for  $M_A\!=\!300$ GeV and three  $(M_S,
\tb)$ sets. $A_t$ is such that $M_h\!=\! 123$--129 GeV and $A_b\!=\!0$.}
\label{approx1}
\vspace*{-7mm}
\end{center}
\end{figure}

Taking into account only the dominant top--stop radiative corrections in the
approximations of eqs.~(A4--A6), Fig.~\ref{approx1} displays the mass of the
heavy CP--even Higgs state (left) and the mixing angle $\alpha$ (right) as a
function of $\mu$ when $\Delta {\cal M}^{2}_{11}$ and $\Delta {\cal
M}^{2}_{12}$ are set to zero (dashed lines) and when they are included (solid
lines). We have assumed the same $(M_S, \tb)$ sets as above and for each value
of $\mu$,  we calculate ``approximate" and `exact"$ M_H$ and $ \alpha$
values assuming $M_h=126\pm 3$ GeV. Even for
large values of the parameter $\mu$ (but $\mu \lsim 3$ TeV), the relative
variation for $M_H$ never  exceeds the $0.5\%$ level while the variation of the
angle $\alpha$ is bounded by $\Delta \alpha \lsim 0.015$. Hence, in this scenario for
the radiative corrections, the approximation of determining the parameters $M_H$
and $\alpha$ from $\tb, M_A$ and the value of $M_h$ is extremely good. We have
again verified that it stays the case when the corrections in the sbottom
sector, with $A_b=A_t$, are included. 

We should note that for higher $M_A$ values, $M_A \gsim 300$ GeV, the
approximation is even better as we are closer to the decoupling limit in which 
one has $M_H\!=\!M_A$ and $\alpha\!=\!\frac{\pi}{2}-\beta$. Lower values,
$M_A \lsim  300$ GeV, are disfavored by the observed $h$ rates ~\cite{R1p,O1}
as  seen later.

In order to check more thoroughly the impact of the subleading corrections
$\Delta {\cal M}^{2}_{11}$, $\Delta {\cal M}^{2}_{12}$, we perform a scan of the
MSSM parameter space using the program {\tt SuSpect} \cite{Suspect} in which the
full two--loop radiative corrections to the Higgs sector  are implemented. For a
chosen  ($\tb$,$M_A$) input set, the soft--SUSY parameters that play an
important role in the Higgs sector are varied in the following ranges:
$|\mu|\leq 3$~TeV,   $|A_t,A_b|\leq 3 M_S$, $1$~TeV$\leq \! M_3 \! \leq \!3$ TeV
and $0.5$~TeV$ \!\leq \!M_S \!\leq  \!3$~TeV ($\approx 3$ TeV is the scale up to
which programs such as {\tt SuSpect} are expected to be reliable). We  assume
the usual relation between the  weak scale gaugino masses  $6 M_1\!= \!3 M_2 \!=
\!M_3$ and set $A_u,A_d, A_\tau\! =\! 0$ (these  last parameters have little
impact). 

We have computed the MSSM Higgs sector parameters all across the parameter space
selecting the points which satisfy the constraint  $123 \! \leq \! M_h \!  \leq
\! 129$ GeV. For each of the points,  we have compared the Higgs parameters  to
those obtained in the simplified MSSM approximation,  $\Delta {\cal M}^{2}_{11}
\! = \! \Delta {\cal M}^{2}_{12} \! =\! 0$, with the lightest Higgs boson mass
as input. We also required ${M}_h$ to lie in the range 123--129 GeV, but allowed
it to be different from the one obtained in  the ``exact" case $\Delta {\cal
M}^{2}_{11}, \Delta {\cal M}^{2}_{12} \neq 0$. 

\begin{figure}[!h]
\begin{center}
\begin{tabular}{cc}
\includegraphics[scale=0.59]{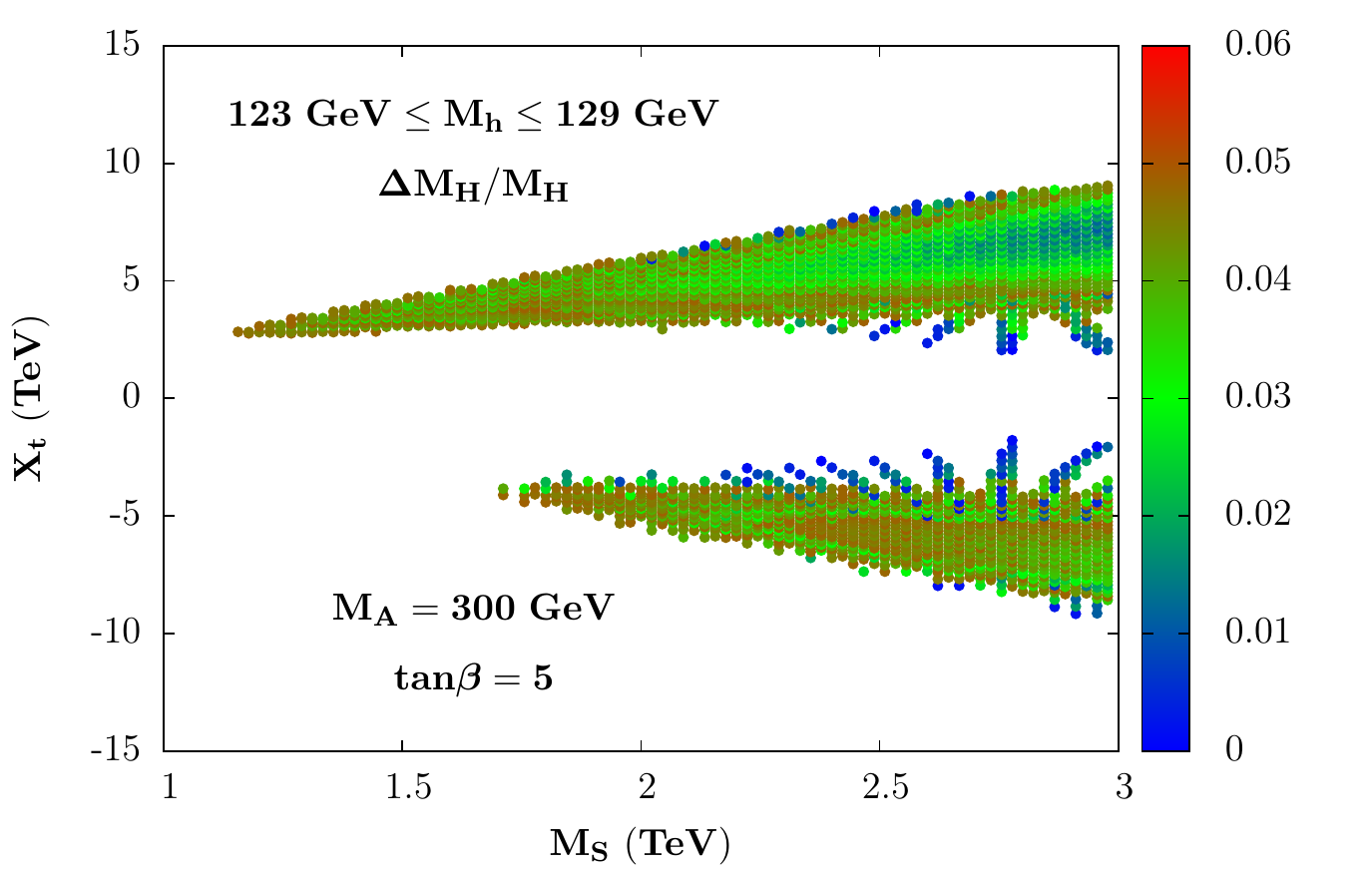}\hspace*{-5mm} &
\includegraphics[scale=0.59]{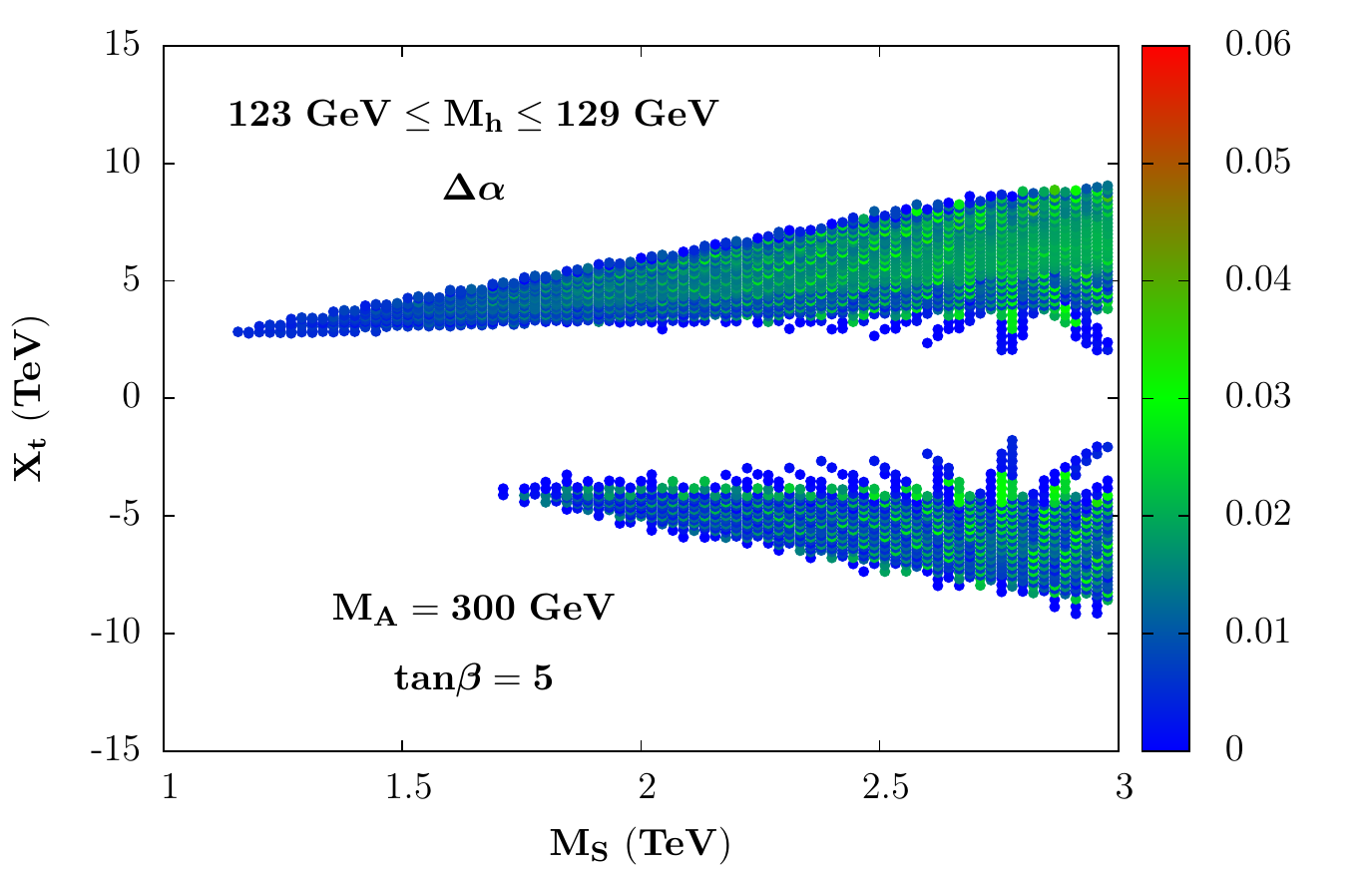} \\
\includegraphics[scale=0.59]{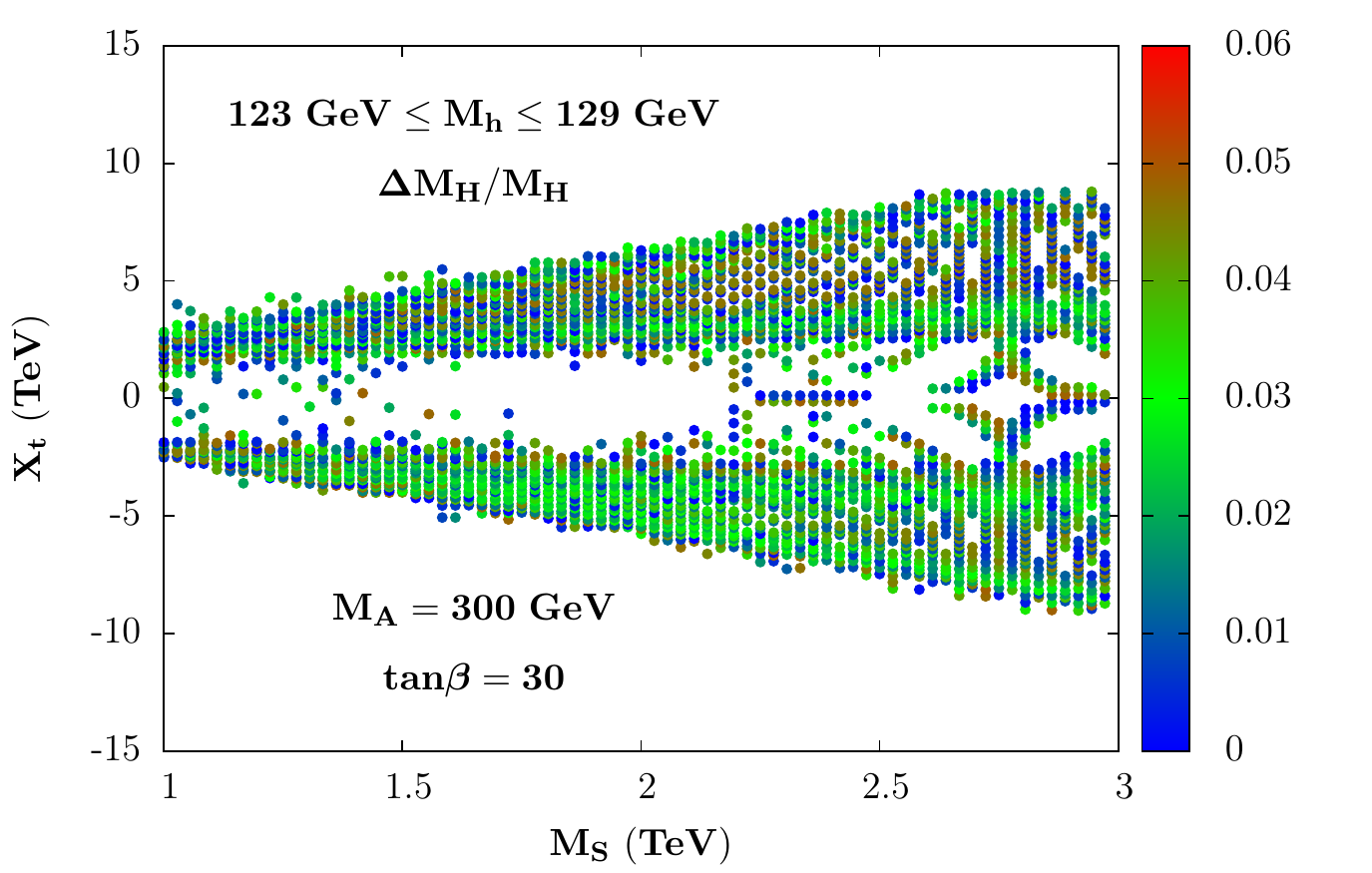}\hspace*{-5mm} &
\includegraphics[scale=0.59]{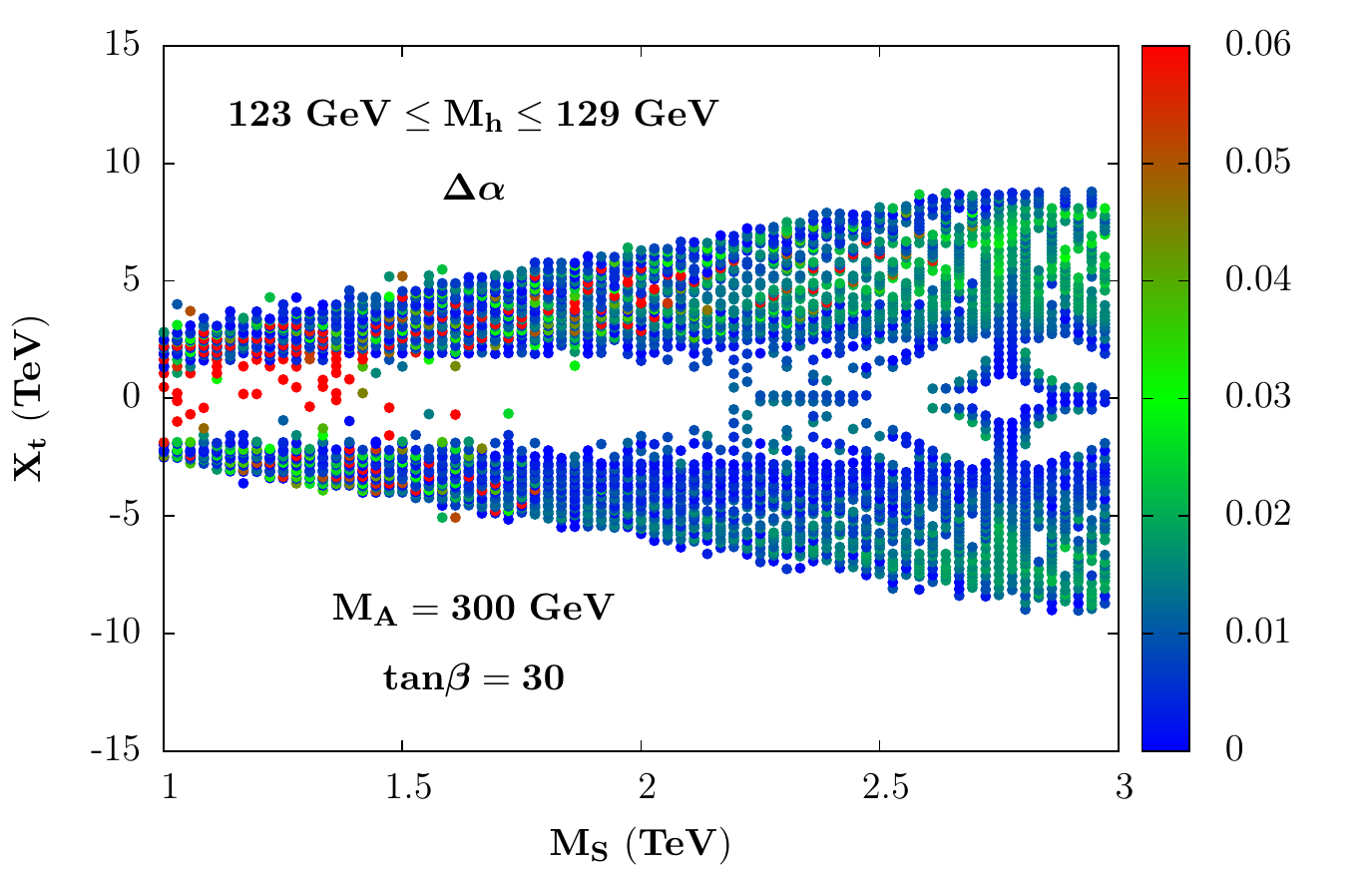} 
\end{tabular}
\caption{\small
The variation of the mass $M_H$ (left) and the mixing angle $\alpha$ (right),
are shown as separate vertical colored scales, in
the plane $[M_{S},X_{t}]$ when the  full two loop corrections are included with
and without the subleading matrix elements $\Delta {\cal M}^{2}_{11}$ and
$\Delta {\cal M}^{2}_{12}$. We take $M_A\!=\!300$ GeV, $\tan\beta=5$ (top) 
and 30 (bottom) and the other parameters are varied as described in the 
text.}
\label{approx2}
\end{center}
\vspace*{-5mm}
\end{figure}

For the mass $M_H$ and the angle $\alpha$,  we display in Fig.~\ref{approx2} the
difference between the values obtained when the two possibilities $ \Delta {\cal
M}^{2}_{11}\!= \!\Delta {\cal M}^{2}_{12} \!= \!0$ and $\Delta {\cal
M}^{2}_{11}, \Delta {\cal M}^{2}_{12} \! \neq \! 0$ are considered. This is
shown  in the plane $[M_{S},X_{t}]$ with $X_t=A_t-\mu \cot \beta$ when all other
parameters are scanned as above. Again, we have fixed the pseudoscalar Higgs
mass to $M_A\!=\!300$ GeV and used the two representative values $\tb=5$ and
$30$.  We have adopted the conservative approach of plotting only points which
maximize these differences.

In all cases, the difference between the two $M_H$ values is very small (in
fact, much smaller than the total decay width $\Gamma_H$), less than a few
percent, while for $\alpha$ the difference does not exceed  $\approx 0.025$ for
low values of $\tb$ but at high $\tb$ values, one can reach the level of
$\approx 0.05$ in some rare situations (large values of $\mu$,  which enhance
the $\mu \tb$ contributions). Nevertheless, at high enough $\tb$,  we are far in
the decoupling regime already for $M_A \gsim 200$ GeV and such a difference does
not significantly affect the couplings of the $h$ and $H$ bosons which,
phenomenologically, are the main  ingredients. 

Hence, even when including the full set of radiative corrections up to two
loops, it is a good approximation to use eqs.~(\ref{wide}) to derive the
parameters $M_H$ and $\alpha$ in terms of the inputs $\tb, M_A$ and the measured
value of $M_h$. In the case of the charged Higgs boson mass, the radiative
corrections are much smaller for large enough $M_A$ and one has, at  the few
percent level (which is again smaller than the total $H^\pm$ decay width),
$M_{H^\pm} \simeq \sqrt { M_A^2 + M_W^2}$ except in very rare
situations\footnote{The physics  of the charged boson, i.e the production and
decay rates,  can be accurately described by $\tb, M_{H^\pm}$ (and  eventually
$\alpha$ if the subleading processes involving the $h$ state are also considered).}
\cite{charged}.

\subsection*{3. Determination of the h boson couplings in a generic MSSM}

A second important issue is the MSSM Higgs couplings. In principle and as
discussed earlier, knowing two parameters such as the pair
of inputs $[\tan\beta, M_A]$ and fixing the value of $M_h$ to its  measured
value, the couplings of the Higgs bosons, in particular $h$,  to fermions and
gauge bosons can be derived, including the generally dominant radiative
corrections that enter in the  MSSM Higgs masses.  Indeed, in terms of the
angles $\beta$ and $\alpha$,   one has for the reduced couplings
 (i.e. normalized to their SM values) of the
lighter $h$ state to third generation $t,b$  fermions and gauge bosons
$V\! = \!W/Z$, 
\begin{eqnarray}  
c_V^0   =   \sin(\beta- \alpha)  \ , \ \  c_t^0   =  
\frac{\cos  \alpha}{\sin\beta} \ , \ \  c_b^0   =   -
\frac{\sin  \alpha}{\cos\beta}  \label{Eq:MSSMlaws} 
\end{eqnarray}
However, outside the regime in which the pseudoscalar $A$ boson and some
supersymmetric particles are very heavy, there are also direct radiative
corrections to the Higgs couplings  not contained in the mass matrix of eq.~(1). 
These can  alter  this simple picture.

First, in the case of $b$--quarks, additional one--loop vertex corrections
modify the tree--level $h b \bar b$ coupling: they grow as $ m_b \mu \tan\beta$
and are thus very large at high  $\tb$. The dominant component comes from the
SUSY--QCD corrections  with sbottom--gluino loops that can be approximated by
$\Delta_b \simeq  2\alpha_s/(3\pi) \times \mu m_{\tilde{g}} \tb /{\rm max}
(m_{\tilde{g}}^2,  m_{\tilde{b}_1}^2,m_{\tilde{b}_2}^2) \label{deltab}$
\cite{Deltab}. 

Outside
the decoupling regime, the $hb\bar b$  coupling receives the
possibly large  correction 
\beq
c_b \approx c_b^0 \times [1- \Delta_b/(1+\Delta_b) \times (1+ \cot\alpha \cot\beta)]  
~~{\rm with}~\tan\alpha \stackrel{M_A \gg M_Z} \to -1/\tb
\label{cb}
\eeq
which would significantly alter the partial width of the decay $h \to b\bar b$ 
that is, in principle, by far the dominant one and, hence, affect the  branching
fractions of  all other decay  modes. 

In addition, the $ht\bar t$ coupling is derived indirectly from the $gg \to h$ 
production cross section and the $h \to \gamma \gamma$ decay branching ratio,
two processes that are generated via triangular loops. In the MSSM, these loops 
involve not only the top quark (and the $W$ boson in the decay $h\to \gamma \gamma$) 
but also contributions from supersymmetric particles, if they are not
too heavy. In the case of the $gg\to h$ process, only the contributions of stops
is generally important. Including the later and working in the limit $M_h
\ll m_t, m_{\tilde{ t_1} }, m_{\tilde{ t_2} }$,  the $hgg$ amplitude can be
(very well) approximated by the expression \cite{Stop}
\beq 
c_t \approx c_t^0 \times  \bigg[ 1 +  \frac {m_t^2}{ 4 m_{\tilde t_1}^2 
m_{\tilde t_2}^2 } ( m_{\tilde t_1}^2 +  m_{\tilde t_2}^2  - (A_t
-\mu\cot\alpha)( A_t+\mu \tan\alpha) \; ) \bigg] 
\label{ct}
\eeq
which shows that indeed, $\tilde t$  contributions can be very large for
sufficiently light stops and in the presence of large stop mixing.  In the  $h
\to \gamma\gamma$ decay rate, because the $t, \tilde t$ electric charges  are
the same, the $ht\bar t$ coupling is shifted by the same amount as  above
\cite{Stop2}. 

If one ignores the usually small $\tilde b$ contributions 
in the $gg\to h$ production and $h\to \gamma \gamma$ decay processes
(in the latter case, it is suppressed  by powers of  the $b$ electric charge
$e_b^2/e_t^2 = \frac14$ in addition) as well as the contributions of other SUSY
particles such as charginos  and stau's in the $h \to \gamma \gamma$ decay 
rate\footnote{The chargino contribution cannot exceed the 10\% level even for
very favorable gaugino-higgsino parameters \cite{Stop2}, while the $\tilde \tau$
contributions  are important  only for extreme values of $\tan\beta$ and $\mu$
\cite{Tau}.},  the leading corrections to the $ht\bar t$ vertex can be
simply accounted for  by using the effective coupling given in eq.~(\ref{ct});
see e.g. Ref.~\cite{R1p}.

Note that in the case of associated production of the $h$ boson with top quarks, 
$gg/q\bar q \to h t\bar t$, it is the parameter $c_t^0$ which should be considered for the direct $ht\bar t$ coupling. 
However, for the time being (and presumably for a long time),  the constraints on the $h$ properties from this 
 process are very weak as the cross section has very large uncertainties. 

One also should note that the couplings of the $h$ boson to $\tau$ leptons and
charm  quarks do not receive the direct corrections of respectively
eqs.~(\ref{cb}) and (\ref{ct}) and one should still have $c_c=c_t^0$ and 
$c_\tau= c_b^0$. However, using $c_{t,b}$ or $c_{t,b}^0$ in this case  has
almost no impact in practice as these couplings appear only in  the branching
ratios for the decays $h \to c\bar c$ and $\tau^+ \tau^-$ which are small, below
5\%, and the direct  corrections cannot be very large (these are radiative
corrections after all). One can thus, in a first approximation, ignore them
and assume that $c_c=c_t$ and $c_\tau=c_b$. Note that BR($h\to c\bar c$)  
cannot be measured at the LHC while the  $h \to \tau^+ \tau^-$ rate is presently 
measured only at the level of $40$\% or so \cite{Barcelone}.

Another caveat is that possible invisible decays (which at present 
are probed directly only for rates that are at the 50\% to 100\% level
\cite{invisible}), can also affect the properties of the observed $h$ particle.
However, a large invisible rate implies that the neutralinos that are considered
as the lightest SUSY particles,   are relatively light and couple significantly
to the $h$ boson,  a situation that is rather unlikely (if the LSP is very
light,  $2m_{\chi_1^0} \lsim M_h$, it should be mostly bino--like and, hence, 
has very suppressed couplings to the Higgs bosons that prefer to couple to
mixtures of higgsinos and gauginos; see for instance Ref.~\cite{Stop2}).   

In the case of large direct corrections, 
the Higgs couplings cannot be described only by the parameters $\beta$ and $\alpha$
as in eq.~(\ref{Eq:MSSMlaws}). One should  
consider at least three independent $h$ couplings, namely 
$c_c=c_t$, $c_\tau=c_b$ and $c_V=c_V^0$ as advocated in Ref.~\cite{R2}. 
This is equivalent to excluding the $h\to \tau \tau$ data from the 
global fit which, in practice, has no significant impact as the experimental
error 
on the signal strength in this channel is presently large. Note that a future
determination of the theoretically clean ratio of the $b\bar b$ and $\tau^+ \tau^-$ signals in $pp\to hV$ gives a direct access to the $\Delta_b$ correction 
outside the decoupling regime \cite{O2}. 

To study the $h$ state at the LHC, we thus define the following
effective Lagrangian,     
\begin{eqnarray} 
{\cal L}_h  & = &  \ c_V \ g_{hWW} \ h \ W_{\mu}^+ W^{- \mu} + \ c_V \ g_{hZZ} \ h \ Z_{\mu}^0 Z^{0 \mu}
\label{Eq:LagEff}\\ &- &   
  c_t \; y_t\;  h  \bar t_L  t_R  -  c_t \; y_c \; h  \bar c_L  c_R  - 
  c_b \; y_b\;   h  \bar b_L b_R  - c_b \; y_\tau \; h  \bar \tau_L \tau_R 
\  + \ {\rm h.c.} \nonumber 
\end{eqnarray}
where $y_{t,c,b,\tau}=m_{t,c,b,\tau}/v$ are the SM Yukawa coupling constants in
the mass eigenbasis ($L/R$ indicates the fermion chirality and we consider  only
the heavy fermions that have substantial couplings to the Higgs boson),  
$g_{hWW} = 2M^2_W/v$ and $g_{hZZ} = M^2_Z/v$ are the electroweak gauge boson
couplings and $v$ is the Higgs vacuum expectation value.

We  present the results for the fits of the Higgs signal strengths in the
various channels
\beq
\mu_X \simeq \sigma( pp \to h) \times {\rm BR}(h \to XX)/  \sigma(
pp \to h)_{\rm SM}  \times {\rm BR}(h \to XX)_{\rm SM}
\eeq
closely following the procedure of Ref.~\cite{O2} but in the case of the 
phenomenological MSSM.   All the  Higgs production/decay channels are considered
and the data used are the latest ones \cite{Barcelone} using the full  $\approx
25~{\rm fb}^{-1}$ statistics for the $\gamma\gamma, ZZ, WW$ channels as well as
the $h\to b\bar b$ and $\tau \tau$ modes for CMS, but only $\approx  17 ~{\rm 
fb}^{-1}$ data for the ATLAS fermionic channels.

\begin{figure}[!t] 
\vspace*{-25mm}
\begin{center}
\begin{tabular}{c}
\includegraphics[width=0.55\textwidth]{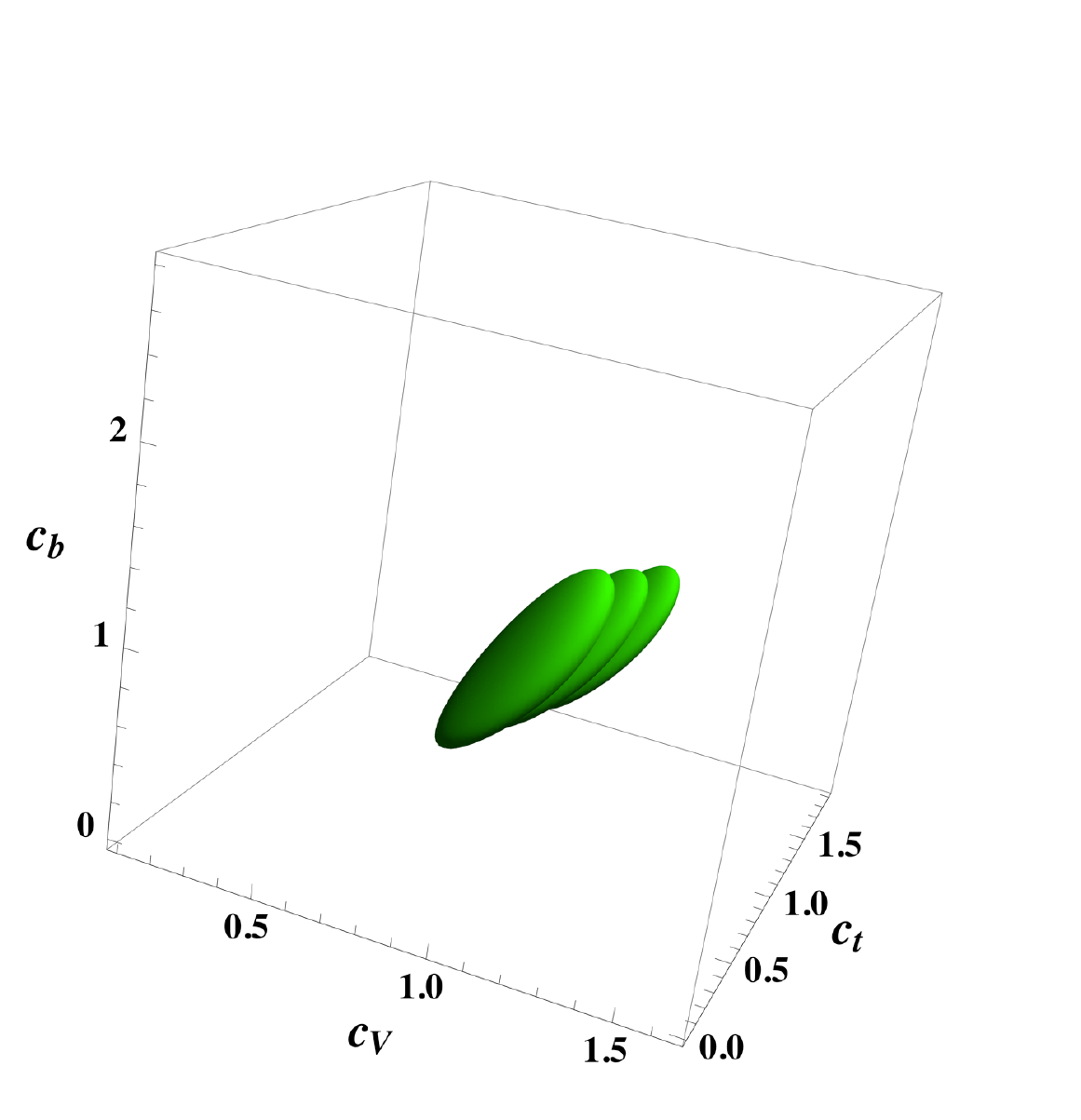}
\includegraphics[width=0.55\textwidth]{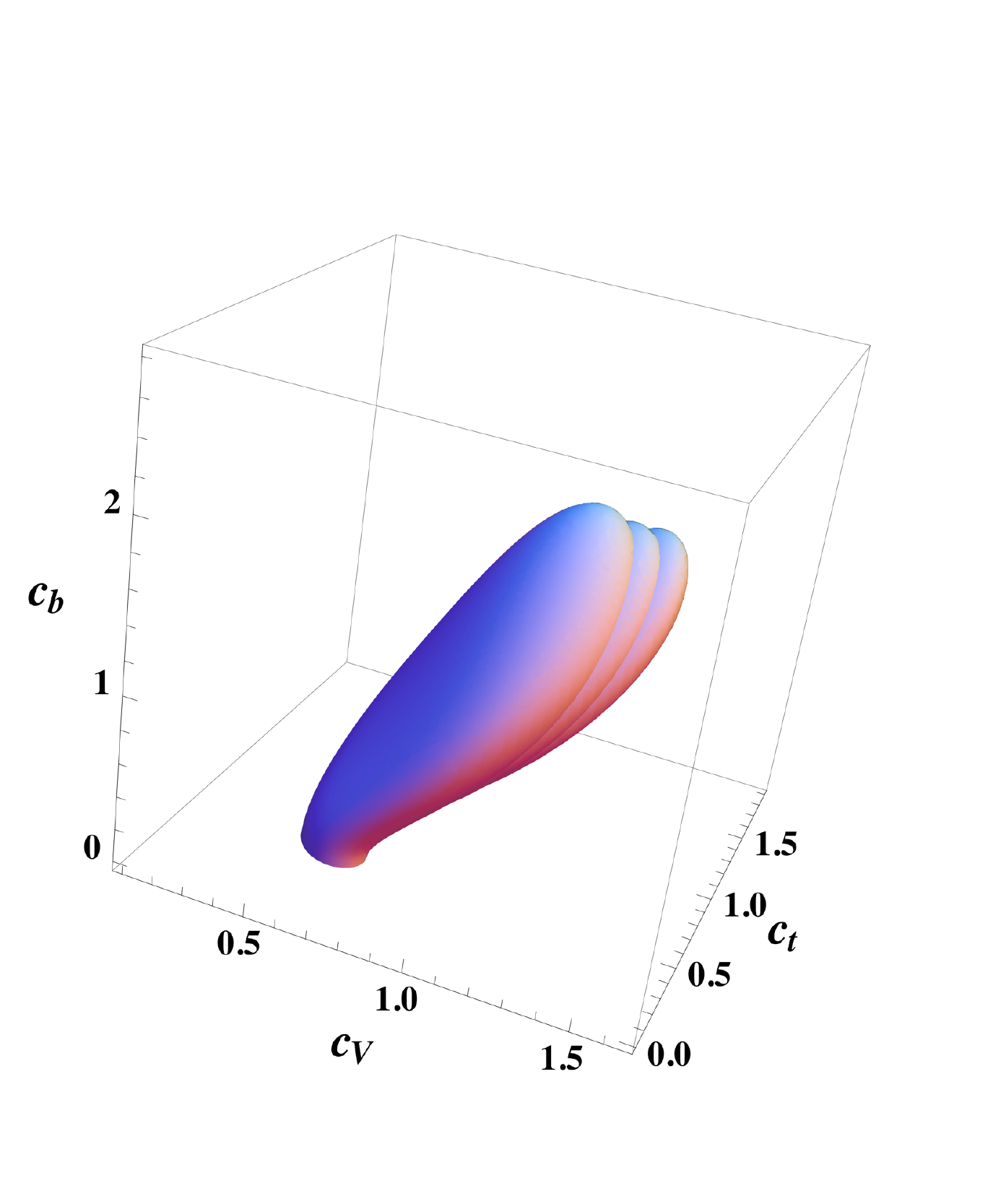}
\end{tabular}
\vspace*{-7mm}
\caption{{\small Best-fit regions at $68\%{\rm CL}$ (green, left) and 
$99\%{\rm CL}$ (light gray, right) for the Higgs signal strengths
in the three--dimensional space $[c_t,c_b,c_V]$. The three overlapped 
regions are associated to central and two extreme  choices of 
the theoretical prediction for the Higgs rates.}}
\label{fig:3D}
\vspace*{-8mm}
\end{center}
\end{figure}

We have performed the appropriate three-parameter fit in the three-dimensional 
space\footnote{Higgs coupling fits have been performed most often in the $[c_V, c_f]$
parameter space with $c_f\! = \! c_t \! = \! c_b \! \dots$. Fits of the LHC data in SUSY scenarios including also the NMSSM
can be found in Ref.~\cite{Sfits} for instance.}
$[c_t, c_b, c_V]$, assuming  $c_c\!=\!c_t$ and $c_\tau\!=\!c_b$  as
discussed above and of course the custodial symmetry relation $c_V\!=\!c_W\!=
\!c_Z$ which holds in supersymmetric models.  The results of this fit are
presented in Fig.~\ref{fig:3D} for $c_t,c_b,c_V \! \geq \! 0$, 
as motivated by the supersymmetric structure of the Higgs couplings (there
is also an exact reflection symmetry under, $c \to -c$ or equivalently $\beta \to \beta + \pi$, leaving the squared
amplitudes of the Higgs rates unaffected).  Again following Ref.~\cite{O2}, we
have treated the theoretical uncertainty as a bias and not as if it were
associated to a statistical distribution and have performed the fit for  values
of the  signal strength~ $\mu_{i} \vert_{ \rm exp} [ 1 \pm \Delta \mu_i/\mu_i
\vert_{\rm th} ]$ with  the theoretical uncertainty $\Delta\mu_i/ \mu_i
\vert_{\rm th}$ conservatively assumed to be $20\%$  for both the gluon and
vector boson fusion mechanisms (because of contamination) and $\approx 5\%$ 
for $h$ production in association with $V=W/Z$ \cite{Baglio}.

The best-fit value for the couplings, when the ATLAS and CMS data are combined, is  
$c_t=0.89, ~ c_b=1.01$ and $c_V=1.02$ with  $\chi^2 =64.8$ ($\chi^2 =66.7$ in the SM).

\begin{figure}[!ht]
\begin{center}
\begin{tabular}{ccc}
\includegraphics[width=0.32\textwidth,height=5cm]{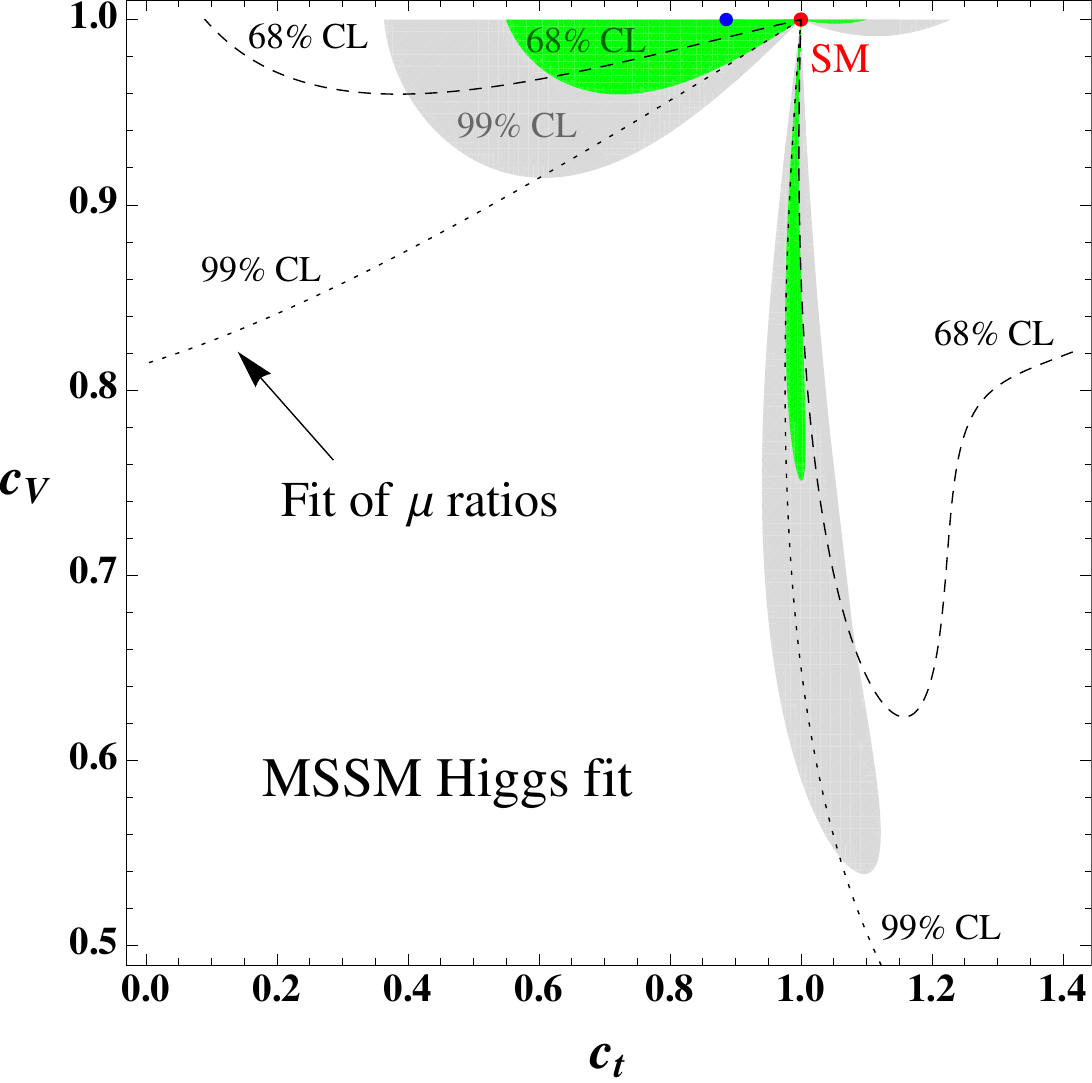} &
\includegraphics[width=0.32\textwidth,height=5cm]{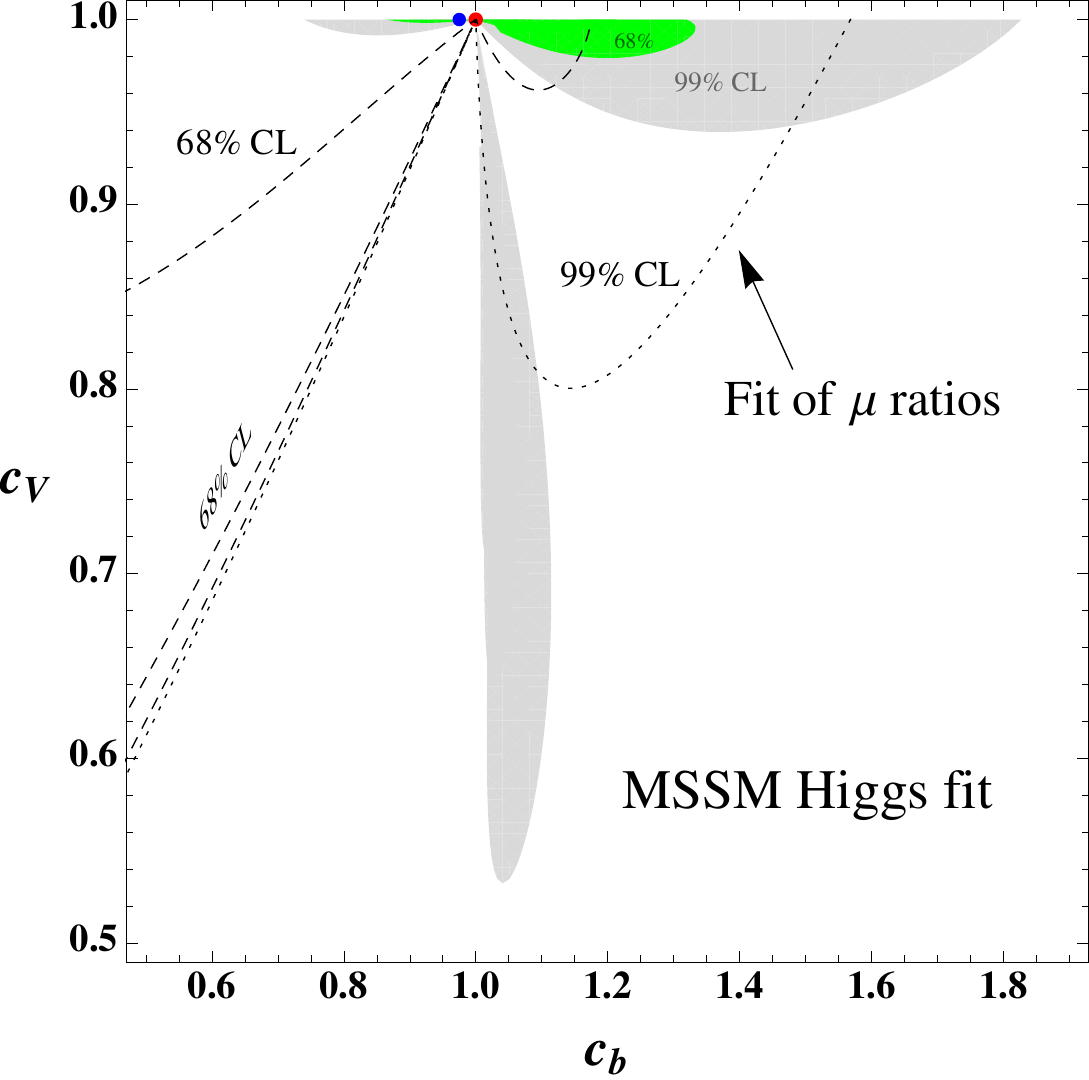} &
\includegraphics[width=0.31\textwidth,height=5cm]{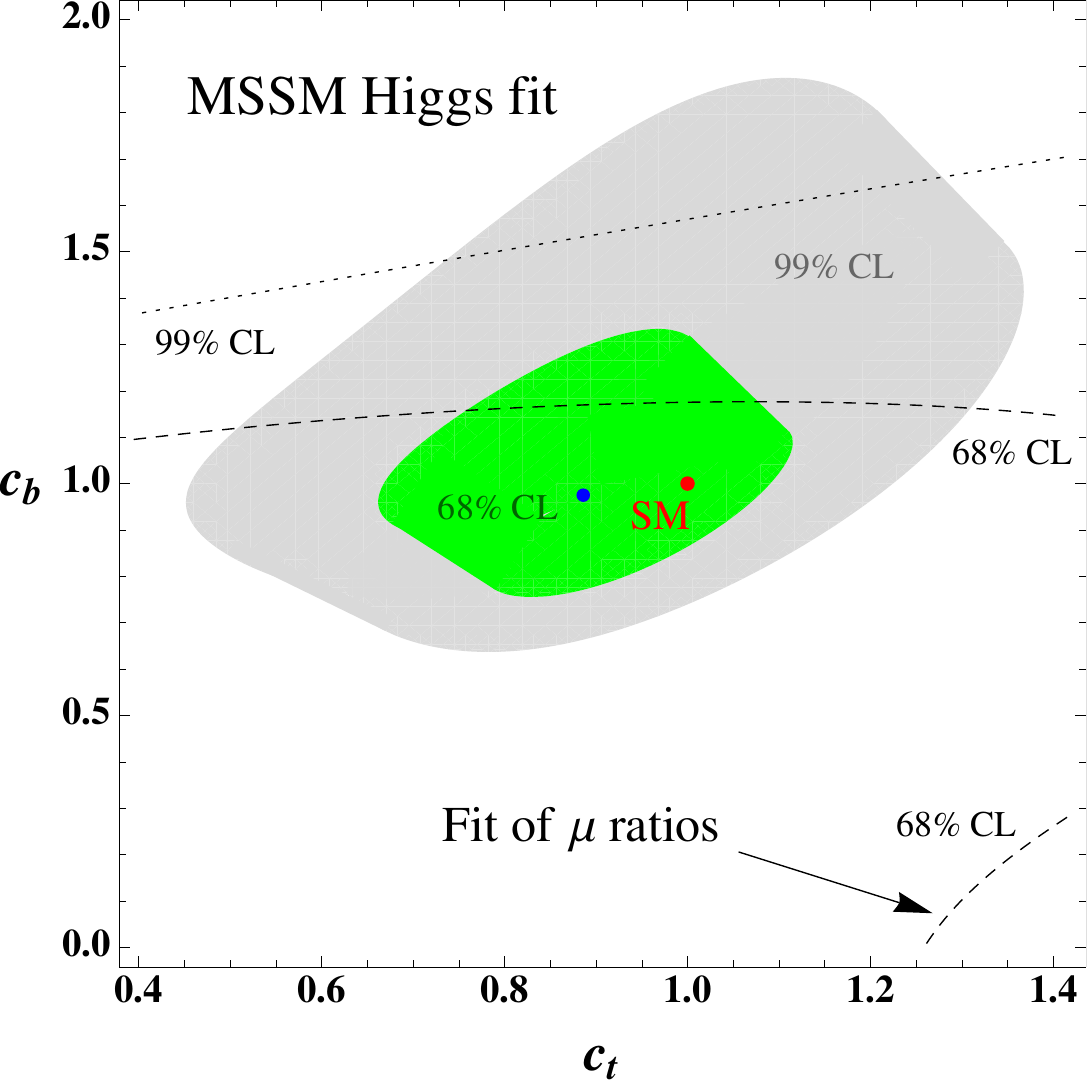}
\end{tabular}
\caption{{\small Best-fit regions at $68\%{\rm CL}$ (green) and  $99\%{\rm
CL}$ (light gray) for the Higgs signal strengths in the  planes $[c_t,c_V]$
(left), $[c_b,c_V]$ (center) and $[c_t,c_b]$ (right).  The theoretical 
uncertainty on the Higgs signal strengths is taken into account as  a bias. The
best-fit contours at $68\%{\rm CL}$ (dashed) and $99\%{\rm CL}$ (dotted)
from the fit of signal strength ratios  are superimposed as well. The SM points are indicated in 
red and the best-fit points in blue.}}
\label{fig:2D}
\end{center}
\vspace*{-5mm}
\end{figure}

In turn, in scenarios where the direct corrections in eqs.~(\ref{cb})-(\ref{ct}) 
are not quantitatively significant (i.e. considering either not too large values 
of $\mu \tan\beta$ or high stop/sbottom masses), 
one can use the MSSM relations of eq.~(\ref{Eq:MSSMlaws}) to reduce the number of effective parameters down to two.
For instance, using  
$c_t   =   \cos \alpha/\sin\beta$ and $c_V   =   \sin(\beta- \alpha)$, one can
derive the following relation, $c_b   \equiv  - \sin \alpha/ \cos\beta =
(1-c_Vc_t)/(c_V-c_t)$.    This allows 
to perform the two-parameter fit in the plane $[c_V,c_t]$.
Similarly, one can study the planes $[c_V,c_b]$ and $[c_t,c_b]$. The two-dimensional fits in these 
three planes are displayed in Fig.~\ref{fig:2D}.
As in the MSSM one has $ \alpha \in [-\pi/2,0]$ and $\tan
\beta \in [1,\sim 50]$, one obtains the following  variation ranges: $c_V \in
[0,1]$, $c_t \in [0,\sqrt{2}]$ and $c_b >0$. 

We also show on these figures the potential constraints obtained from fitting
ratios of the Higgs signal strengths (essentially the two ratios $R_{\gamma
\gamma}= \mu_{\gamma \gamma}/\mu_{ZZ}$ and $R_{\tau \tau}=
\mu_{\tau\tau}/\mu_{WW}$)  that are not or much less affected by the QCD 
uncertainties  at the production level~\cite{O2}. In this two--dimensional case,
the best-fit points   are located at $(c_t=0.88$, $c_V=1.0)$, $(c_b=0.97$,
$c_V=1.0)$ and ($c_t=0.88$, $c_b=0.97$). Note that although  for the best--fit
point one has $c_b \lsim 1$, actually $c_b \gsim 1$ in most of the  $1\sigma$ region. 

\begin{figure}[!h]
\begin{center}
\begin{tabular}{c}
\includegraphics[width=0.45\textwidth]{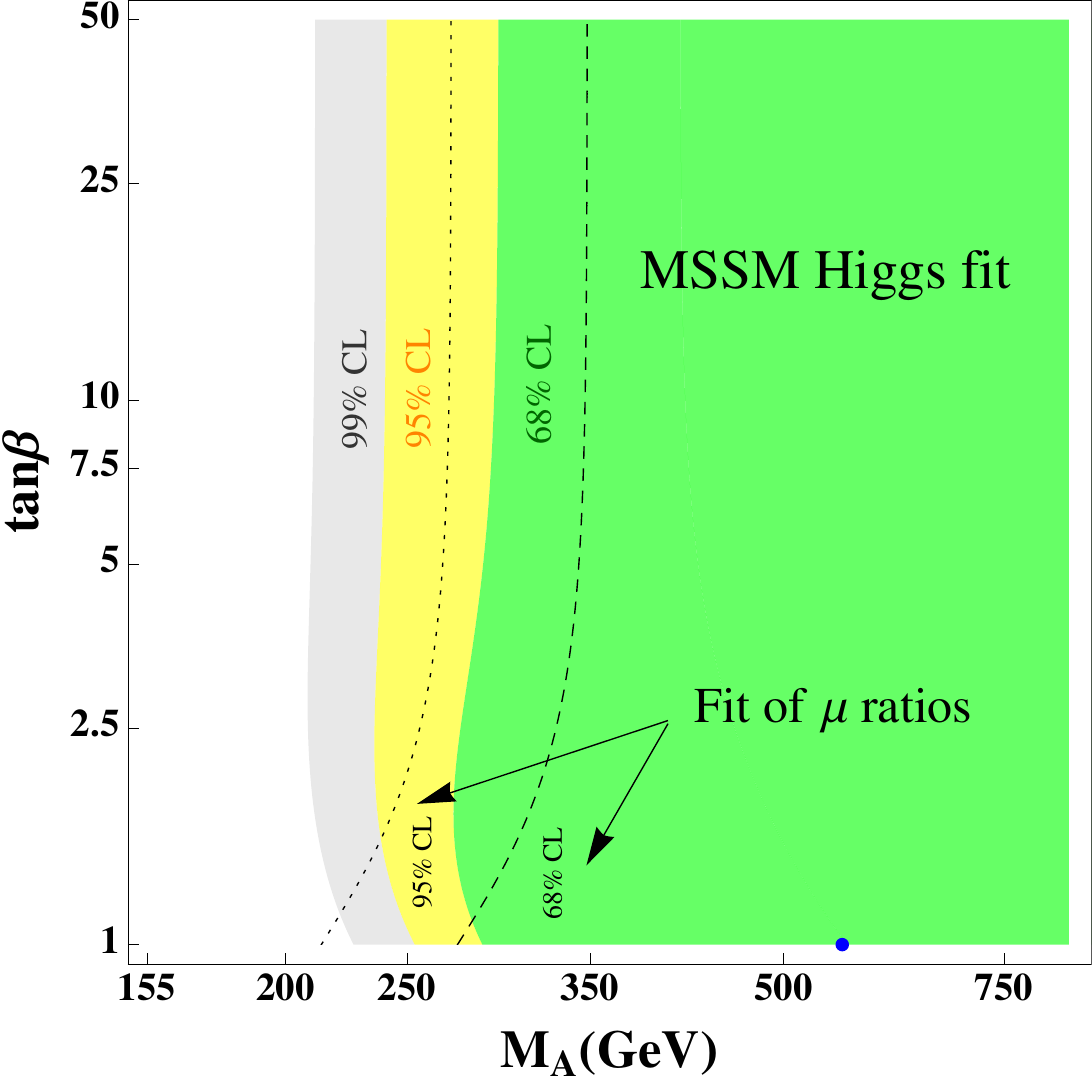}~~
\includegraphics[width=0.45\textwidth]{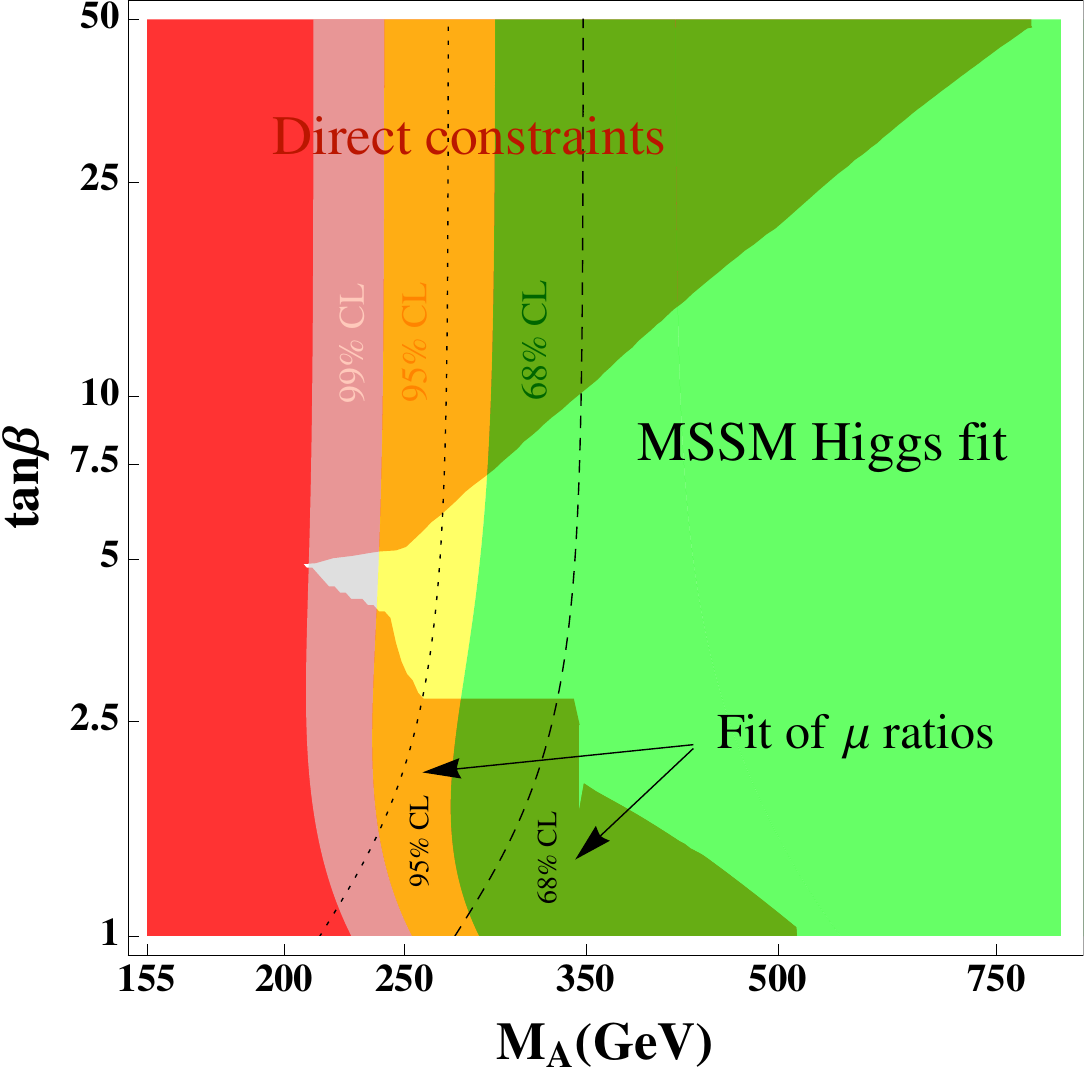}
\end{tabular}
\caption{{\small Left: best-fit regions at $68\%{\rm CL}$ (green), $95\%{\rm
CL}$ (yellow)  and $99\%{\rm CL}$ (light gray) for the Higgs signal strengths in
the plane $[\tan \beta, M_A]$; the best--fit point is shown in blue and   the
theoretical uncertainty  is taken into account as a bias   as in the previous
figures. The best-fit contours at $1\sigma$ (dashed) and  $2\sigma$
(dotted) for the signal strength ratios are also shown. Right:  we
superimpose on these constraints the excluded regions (in red, and as a shadow
when superimposed on the best-fit regions) from the direct searches of the
heavier Higgs bosons at the LHC following the analysis of Ref.~\cite{O1}.}}

\label{fig:2DMA}
\vspace*{-5mm}
\end{center}
\end{figure}

Alternatively, using the expressions of eq.~(\ref{Eq:MSSMlaws}), one can also
realize a two-parameter fit in the $[\tan \beta,  \alpha]$ plane\footnote{This
corresponds in fact to the case of a two--Higgs doublet model in which the
direct corrections are expected to be small in contrast to the SUSY case: one
can then parametrise the couplings of the $h$ boson,  that are given by
eq.~(\ref{Eq:MSSMlaws}), by still two parameters $\alpha$ and $\beta$ but with
the angle $\alpha$ being a free input.}. However, using the expressions of
eq.~(\ref{wide}) for the mixing angle $ \alpha$ and fixing $M_h$ to the measured
value $M_h \! \approx 125$ GeV, one can perform a fit in the plane $[\tan
\beta,  M_A]$. This is shown in the left--hand side of Fig.~\ref{fig:2DMA} where
the 68\%CL, 95\%CL and  99\%CL contours from the signal strengths only are
displayed when, again,  the theoretical uncertainty is considered as a bias. We
also  display the best-fit contours for the signal strength ratios at the
68\%CL  and 95\%CL.  The best-fit point for the signal strengths when the
theoretical uncertainty is set to zero, is obtained  for the  values
$\tan\beta\!=\! 1$  and $M_A \! = \! 557 \; {\rm GeV}$,   which implies
for the other parameters, when the radiative corrections entering the Higgs
masses and the angle $ \alpha$ are derived using the information $M_h=125$~GeV~:
$M_H= 580$~GeV, $M_{H^\pm}= 563$~GeV and $ \alpha=-0.837~{\rm rad}$. Regarding
this best-fit point, one should note that the $\chi^2$ value is relatively
stable all over the $1\sigma$ region shown in Fig.~\ref{fig:2DMA}.

It is interesting to superimpose on these indirect limits in the $[\tan
\beta,M_A]$ plane, the direct constraints on the heavy $H/A/H^\pm$ boson
searches performed by the ATLAS  and CMS collaborations as shown 
in the right--hand side of Fig.~\ref{fig:2DMA}. As discussed in
Ref.~\cite{O1} (see also Ref.~\cite{ABM}), besides the limits from the 
$A/H \to \tau^+ \tau^-$ and to a
lesser extent $t\to b H^+  \to b \tau \nu$ searches which exclude high $\tb$
values and which can be extended to very low $\tb$ as well, there are also
limits  from adapting to the MSSM the high mass SM Higgs searches  in  the
channels\footnote{At low $\tb$,  channels such as $A\to hZ$ and  
$H \to hh$ need also to be considered \cite{O1}. In the latter case, special care is 
needed in the treatment of the trilinear $Hhh$ coupling as will be discussed 
in Ref.~\cite{preparation}.} $H\to WW$ and $ZZ$ as well as
the searches for heavy resonances decaying into
$t\bar t$ final states that exclude low values of $\tb$ and $M_A$. For values
$250 \lsim M_A \lsim  350$ GeV, only the intermediate $\tan\beta \approx 2$--10 
range is still allowed.

\subsection*{4. Conclusion} 

We have discussed the hMSSM, i.e. the MSSM that we seem to have after the
discovery of the Higgs  boson at  the LHC that we identify with the lighter $h$
state. The mass  $M_h \approx 125$ GeV and the non--observation of SUSY 
particles, seems to indicate that the soft--SUSY breaking scale might be large,
$M_S \gsim 1$ TeV. We have shown, using both approximate analytical formulae and
a scan of the MSSM parameters, that the MSSM Higgs sector can be described to a
good approximation by only the two parameters $\tb$ and  $M_A$ if the
information $M_h \! = \! 125$ GeV is used. One could then ignore the radiative
corrections to the Higgs masses and their complicated dependence  on the MSSM
parameters and use a simple formula to derive the other parameters of the
Higgs sector, $\alpha$,  $M_H$ and $M_{H^\pm}$. 

In a second step, we have shown that to  describe accurately the $h$ 
properties when the direct radiative corrections are also important, 
the three couplings $c_t, c_b$ and $c_V$ are needed besides the $h$ mass. 
We have performed a fit of these couplings using the latest LHC data
and taking into account properly the theoretical uncertainties. 
In the limit of heavy sparticles (i.e. with small direct corrections), the best fit point 
turns out to be at low $\tb$, $\tb \! \approx \! 1$,  and with a not too high 
CP--odd Higgs mass, $M_A \approx 560$ GeV. 

\begin{figure}[!h]
\vspace*{-2mm}
\begin{center}
\begin{tabular}{c}
\includegraphics[width=0.5\textwidth]{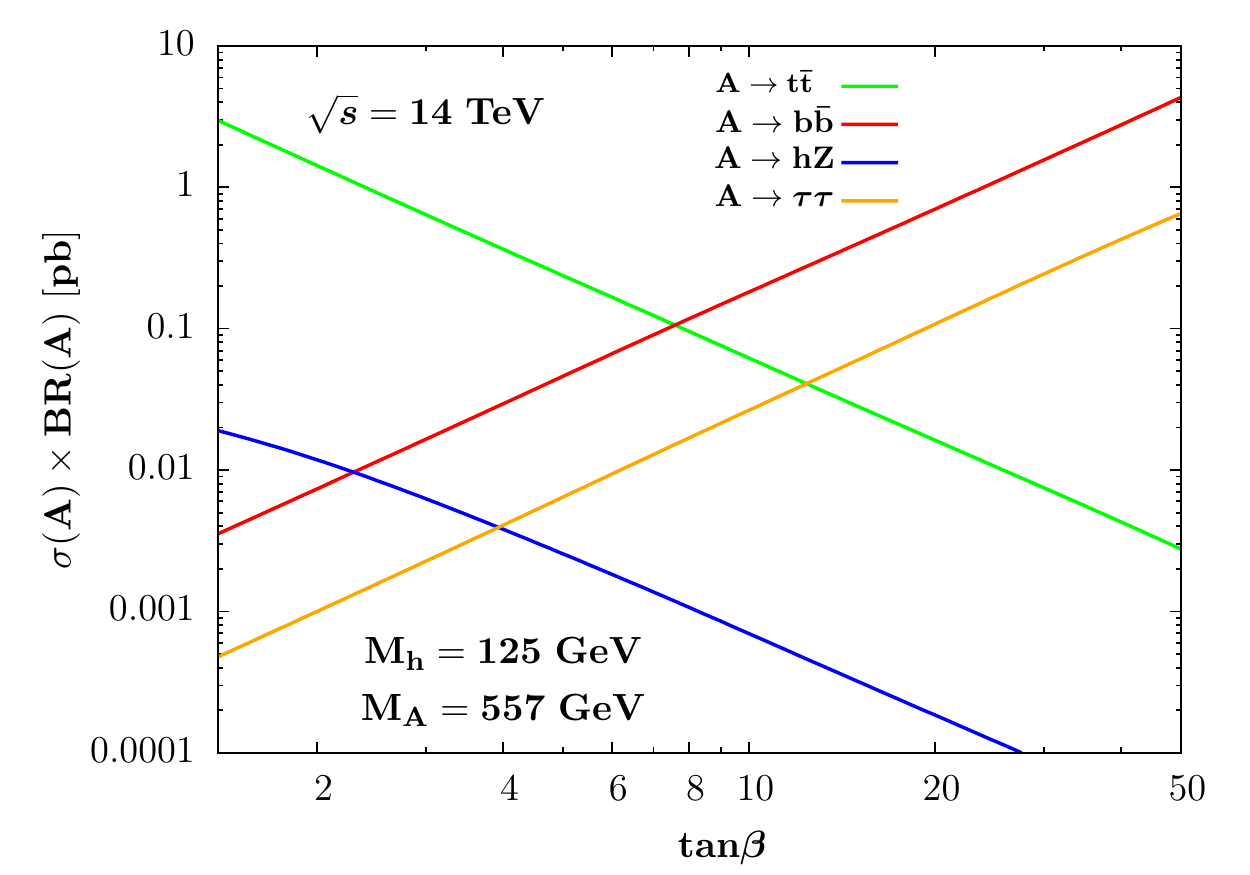}
\includegraphics[width=0.5\textwidth]{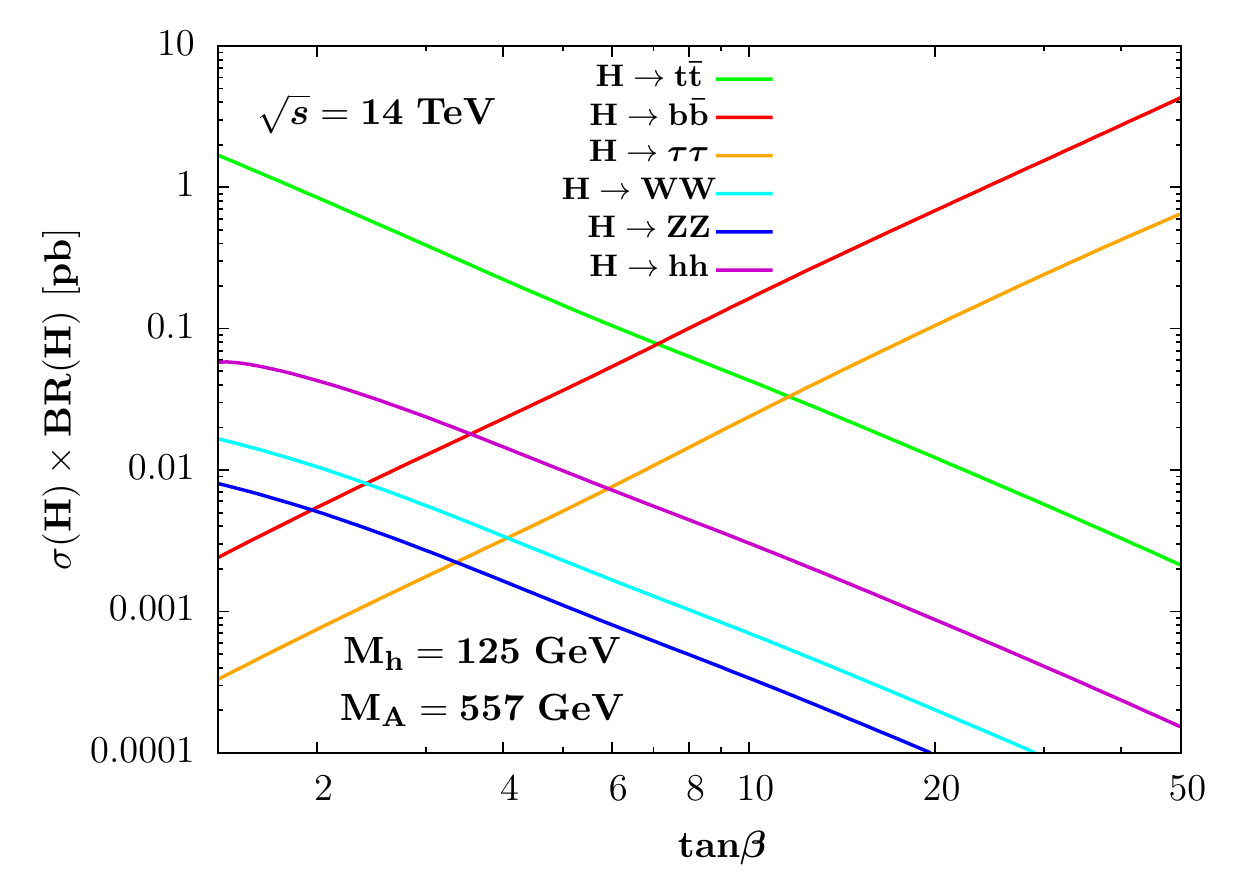}
\end{tabular}
\vspace*{-4mm}
\caption{{\small The cross section times branching fractions for the $A$ (left)
and $H$ (right) MSSM Higgs bosons at the LHC with $\sqrt s=14$ TeV as a function
of $\tb$ for the best--fit mass $M_A=557$ GeV and with $M_h=125$ GeV. For the production,
we have taken into account only the gluon and bottom quark fusion processes and 
followed the analysis given in Ref.~\cite{O1}.}}
\label{LHC14}
\vspace*{-5mm}
\end{center}
\end{figure}

The phenomenology of this particular point is quite interesting. 
First, the heavier Higgs particles will be accessible in the next LHC 
run at least in the channels  $A,H \to t \bar t$ and presumably also in
the modes $H\to WW, ZZ$ as the rates are rather large for $\tb \approx 1$. 
This is shown in Fig.~\ref{LHC14} where the cross sections  times decay 
branching ratios  for $A$ and $H$ are displayed as a function 
of $\tb$ for the choice $M_A=557$ GeV for $\sqrt s=14$ TeV. Further more, 
the correct relic abundance of the LSP neutralino
can be easily obtained through  $\chi_1^0 \chi_1^0 \to A \to t\bar t$ 
annihilation by allowing the parameters $\mu$ and $M_1$ to be 
comparable and have an LSP mass close to
the $A$--pole, $m_{\chi_1^0} \approx \frac12 M_A$. The SUSY
spectrum  of this low $\tb$ region will be discussed in more detail 
in a separate publication \cite{preparation}.\smallskip

\noindent {\bf Acknowledgements}:  
We thank Pietro Slavich for discussions.
AD and LM thank the CERN theory division for the kind hospitality offered to 
them. AD is supported by the ERC Advanced Grant Higgs@LHC and   
GM by the Institut Universitaire de France (IUF).\newpage

\subsection*{Appendix: approximating the radiative corrections}
\setcounter{equation}{0}
\renewcommand{\theequation}{A\arabic{equation}}
  
The radiative corrections to the  CP--even Higgs boson mass matrix 
can be written as
\begin{eqnarray}
{\cal M}^2 = \left[ \begin{array}{cc} {\cal M}_{11}^2 + \Delta {\cal M}_{11}^2
& {\cal M}_{12}^2 + \Delta {\cal M}_{12}^2 \\ 
  {\cal M}_{12}^2 + \Delta {\cal M}_{12}^2 
&  {\cal M}_{22}^2 + \Delta {\cal M}_{22}^2  
\end{array} \right]
\label{HmatrixRC}
\end{eqnarray}
The leading one--loop radiative corrections $\Delta {\cal M}_{ij}^2$ to the 
mass matrix are controlled by the top Yukawa coupling $\lambda_t =  m_t/v
\sin\beta$ which appears with the fourth power. One can obtain a very simple 
analytical expression if only this contribution is taken into account 
\cite{CR-1loop}
\beq
\Delta {\cal M}_{11}^2& \sim & \Delta {\cal M}_{12}^2 \sim 0 \ , \nonumber \\
\Delta {\cal M}_{22}^2& \sim & \epsilon = \frac{3\, \bar{m}_t^4}{2\pi^2 v^2\sin^
2\beta} \left[ \log \frac{M_S^2}{\bar{m}_t^2} + \frac{X_t^2}{\,M_S^2} \left( 1 -
\frac{X_t^2}{12\,M_S^2} \right) \right]
\label{higgscorr}
\eeq
where $M_S$ is the geometric average of the stop masses $M_S =\sqrt{
m_{\tilde{t}_1}m_{\tilde{t}_2}} $, $X_{t}$ is the stop mixing parameter given by
$X_t= A_t- \mu/\tb$ and  $\bar{m}_t$ is the running ${\rm \overline{MS}}$ top
quark mass to account for the leading two--loop QCD  corrections in a
renormalisation--group improvement. \s 

A better approximation, with some more renormalisation--group improved  
two--loop QCD and electroweak corrections included is given by \cite{CR-eff}
\begin{eqnarray} 
\Delta {\cal M}_{22}^2 = \frac{3}{2\pi^2}\frac{m_t^4}{v^2 \sin^2\beta}\left[ \frac{1}{2}
\tilde{X}_t + \ell_S
+\frac{1}{16\pi^2}\left(\frac{3}{2}\frac{m_t^2}{v^2}-32\pi\alpha_s
\right)\left(\tilde{X}_t \ell_S+ \ell_S^2\right) \right]\,, \label{mhsm}
\end{eqnarray}
where  $\ell_S= \log (M_S^2/m_t^2)$ and  using $x_t  =  X_t/M_S=(A_t-\mu
\cot\beta)/M_S$ one has  $\tilde{X}_{t} = 2 x_t^2 (1 - x_t^2/12)$  with $A_t$ 
the trilinear Higgs-stop coupling and $\mu$  the higgsino mass parameter.

Other soft SUSY--breaking parameters, in particular $\mu$ and $A_b$
(and in general the corrections controlled by the bottom Yukawa coupling 
$\lambda_b =  m_b/v \cos\beta$ which at large value of the product $\mu
\tb$, provide a non--negligible correction to ${\cal M}_{ij}^2$) can also have an
impact on the loop corrections.  Including these subleading contributions at
one--loop, plus the leading logarithmic contributions at two--loops, the
radiative corrections to the CP--even mass matrix elements can still be written
in a compact form \cite{Mh-max}
\beq
\Delta {\cal M}_{11}^2 &=& - \frac{v^2 \sin^2\beta}{32 \pi^2} \bar{\mu}^2 
\bigg[  x_t^2 \lambda_t^4 (1+ c_{11} \ell_S) + a_b^2 \lambda_b^4  
(1+ c_{12} \ell_S) \bigg] \nonumber \\
\label{RG:approximation}
\Delta {\cal M}_{12}^2 &=&  - \frac{v^2 \sin^2\beta}{32 \pi^2} \bar{\mu} 
\bigg[  x_t \lambda_t^4 (6- x_t a_t) (1+ c_{31} \ell_S) - \bar \mu^2 a_b 
\lambda_b^4  (1+ c_{32} \ell_S) \bigg] \\
\Delta {\cal M}_{22}^2 &=&  \frac{v^2 \sin^2\beta}{32 \pi^2} 
\bigg[ 6 \lambda_t^4 \ell_S (2+ c_{21} \ell_S) + 
x_t a_t  \lambda_t^4  (12 - x_t a_t) (1+ c_{21} \ell_S) 
 - \bar \mu^4 \lambda_b^4  (1+ c_{22} \ell_S) \bigg] \nonumber  
\eeq
where the additional abbreviations $\bar \mu=\mu/M_S$ and 
$a_{t,b}= A_{t,b}/M_S$  have been used.  The factors $c_{ij}$
take into account the leading two--loop corrections due to the top and bottom 
Yukawa couplings and to the strong coupling constant $g_s=\sqrt{4\pi \alpha_s}$; they read
\beq
c_{ij}= \frac{1}{32\pi^2} (t_{ij}\lambda_{t}^2+ b_{ij}\lambda_{b}^2 -32 g_s^2) 
\eeq
with the various coefficients given by
$(t_{11}, t_{12}, t_{21}, t_{22}, t_{31}, t_{32})  =  (12,-4,6,-10,9,7)$ and
$(b_{11}, b_{12}, b_{21}, b_{22}, b_{31}, b_{32})= (-4,12,2,18,-1,15)$. 

The expressions eq.~(\ref{RG:approximation}) provide a good approximation of 
the bulk of the radiative corrections \cite{Mh-max}. However, one needs to 
include the full  set of corrections  to have precise predictions for the  Higgs
boson masses and couplings as discussed at the end of section 2.

\begin{small}

\end{small}
\end{document}